\documentclass[aps,prl,twocolumn,superscriptaddress,showpacs]{revtex4-1}
\usepackage{amsmath}
\usepackage{amssymb}
\usepackage{braket}
\usepackage{graphicx}
\usepackage{hyperref}

\begin{document}
    \title{Detection of Nonlocal Spin Entanglement\\ by Light Emission from a
        Superconducting p-n Junction}

    \author{Alexander Schroer}
    \affiliation{Institut f\"ur Mathematische Physik, Technische Universit\"at
        Braunschweig, D-38106 Braunschweig, Germany}
    \author{Patrik Recher}
    \affiliation{Institut f\"ur Mathematische Physik, Technische Universit\"at
        Braunschweig, D-38106 Braunschweig, Germany}
    \affiliation{Laboratory for Emerging Nanometrology Braunschweig, D-38106 
        Braunschweig, Germany}

    \begin{abstract}
        We model a superconducting p-n junction in which the n- and the p-sides
        are contacted through two optical quantum dots (QDs), each embedded into
        a photonic nanocavity. Whenever a Cooper pair is transferred from the
        n-side to the p-side, two photons are emitted. When the two electrons of
        a Cooper pair are transported through different QDs,
        polarization-entangled photons are created, provided that the Cooper
        pairs retain their spin singlet character while being spatially
        separated on the two QDs. We show that a CHSH Bell-type measurement is
        able to detect the entanglement of the photons over a broad range of
        microscopic parameters, even in the presence of parasitic processes and
        imperfections. 
    \end{abstract}

    %73.40.Lq 	Other semiconductor-to-semiconductor contacts, p-n junctions, 
    %           and heterojunctions
    %74.45.+c 	Proximity effects; Andreev reflection; SN and SNS junctions
    %42.50.Pq 	Cavity quantum electrodynamics; micromasers
    %03.65.Ud 	Entanglement and quantum nonlocality (e.g. EPR paradox, Bell's 
    %           inequalities, GHZ states, etc.) (for entanglement production and
    %           manipulation, see 03.67.Bg; for entanglement measures, witnesses
    %           etc., see 03.67.Mn; for entanglement in Bose-Einstein
    %           condensates, see 03.75.Gg)

    \maketitle

    Quantum entanglement is a resource for quantum computers and quantum
    communication protocols \cite{nielsen00}. For massive particles like
    electrons, spin entanglement could be created exploiting superconductors
    (SCs) \cite{choi00} and crossed Andreev reflection (CAR)
    \cite{torres99,falci01}. Splitting a Cooper pair (CP) into spatially
    separated normal leads has been investigated in detail theoretically
    \cite{recher01,lesovik01,recher02,bena02,recher03,yeyati07,
    cayssol08,sato10} and realized experimentally \cite{hofstetter09,herrmann10,
    das12,wei10}. However, detecting the spin entanglement of the separated
    electrons remains an open challenge. Different detection schemes have been
    proposed which are either based on the violation of a Bell inequality using
    current cross-correlation (noise) measurements \cite{kawabata01,
    chtchelkatchev02,samuelsson03,beenakker03,samuelsson04-1, sauret05,chen12},
    spin-filtered current measurements \cite{braunecker13}, or exploiting the
    statistical \cite{burkard00,egues02,burkard03,hu04,samuelsson04-2,egues05,
    giovannetti06,prada06,mazza13} and many-body \cite{schroer14} properties of
    electronic beam splitters. All of these proposals, however, are much more
    involved than the detection of polarization-entangled optical photons, which
    has been demonstrated successfully \cite{freedman72,aspect82,weihs98} by
    violating a Bell inequality \cite{bell64,clauser69}. Quite naturally, the
    question arises if superconductivity and optics can be combined. It has been
    discussed that p-n junctions in contact with superconductors can exhibit
    enhanced radiation intensity \cite{hanamura02,asano09} with experimental
    demonstration \cite{sasakura11}, pairwise (entangled) emission of light
    \cite{suemune06,recher10,hassler10,baireuther14,hayat14}, Josephson
    radiation at optical frequencies \cite{recher10}, squeezed light
    \cite{baireuther14} and laser effects \cite{godschalk11,godschalk13,
    godschalk14}.  Here, we would like to go a step further and ask the question
    whether nonlocal spin entanglement can be mapped to optical photons in a
    superconducting p-n junction. Existing theoretical proposals use the optical
    recombination of electron and hole singlets in a tunnel-coupled double
    quantum dot (QD) \cite{gywat02}, transfer of spin-entangled electrons into
    two empty optical quantum dots \cite{cerletti05}, and the simultaneous
    transport of spin-entangled electrons and holes into two optical quantum
    dots embedded in nanocavities \cite{budich09}. Very recently, the conversion
    of CPs into photons by laser excitation \cite{nigg14} has also been studied.

     \begin{figure}
        \includegraphics{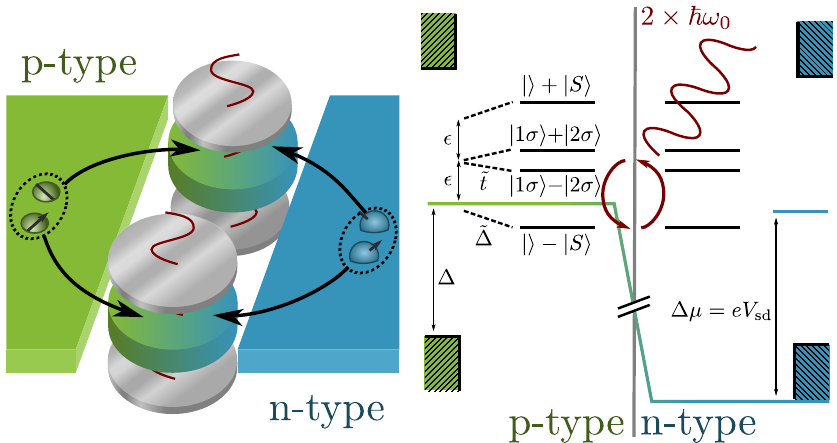}
        \caption{Left: two QDs (1,2) are tunnel-coupled to superconducting 
            p-type and n-type leads. With a sufficiently large Coulomb repulsion
            on the QDs, the electrons or holes of incoming CPs are split, such
            that each QD contains one electron and one hole. Upon recombination
            polarization-entangled photons are emitted into two nanocavities
            (gray) with resonance frequency $\omega_0$ and detected by an
            optical Bell test. Right: Lowest relevant energy levels of the n and
            p double QD systems. The SC leads hybridize the empty state and the
            singlet state, $\Ket{}\pm\Ket{S}$, and the singly occupied states,
            $\Ket{1\sigma}\pm\Ket{2\sigma}$, creating a closed two-photon
            emission cycle.}
        \label{fig-setup}
    \end{figure}
    
    In this Letter, we investigate a superconducting p-n junction, where n- and
    p-sides are contacted via a double QD structure offresonantly embedded into
    two photonic nanocavities (Fig.~\ref{fig-setup}). Due to the superconducting
    pairing, the QDs are populated with spin-entangled hole pairs and
    spin-entangled electron pairs, which, upon recombination, emit pairs of
    polarization-entangled optical photons. Processes which leave behind
    quasiparticle excitations in the QDs or in the SC leads are suppressed by a
    sufficiently narrow spectral width of the cavities and thus photons are
    produced only in pairs with a simultaneous transport of a CP through the
    device. We show that the photons emitted from the cavities violate a CHSH
    Bell inequality when CAR is finite, i.e., when the two electrons or holes of
    a CP retain their spin entanglement when split. We take into account
    parasitic processes in which two photons are emitted into the same cavity
    [arising either from sequential emission after elastic cotunneling (ECT)
    between the QDs or from imperfect splitting], and show within a microscopic
    model that entanglement can be detected nevertheless over a broad range of
    realistic parameters.
  
    \emph{Photonic model.}---%
    We start with an intuitive effective model for the photons in the two
    cavities that will be derived formally later on, 
    \begin{align}
        H_\text{ph}=\sum_{i\xi}\hbar\omega&a^\dagger_{i\xi}a_{i\xi}
            +t_\text{ph} a^\dagger_{i\xi}a_{\bar i\xi} \notag\\
            &+\Bigl(\frac{\Delta_\text{ph}}{2}a_{i\xi}a_{\bar i\bar\xi}
            +\frac{\Lambda_\text{ph}}{2} a_{i\xi}a_{i\bar\xi}
            +\text{h.c.}\Bigr),
        \label{eq-hphoton}
    \end{align}
    where $a_{i\xi}$ annihilates a photon with circular polarization
    $\xi=R,L$ in cavity $i=1,2$ with mode energy $\hbar\omega$ (counted from the
    source-drain bias $eV_\text{sd}$). Because the cavities are coupled to the
    superconducting QDs, there is a finite amplitude $\Delta_\text{ph}$ to
    inject a {\it nonlocal} photon pair into different cavities $i$ and $\bar i$
    and an amplitude $\Lambda_\text{ph}$ to inject a pair {\it locally} into the
    same cavity. Because the pairs can be traced back to one electron singlet
    and one hole singlet, they always come with opposite polarizations $\xi$ and
    $\bar\xi$. We can choose $\Delta_\text{ph}$ and $\Lambda_\text{ph}$ real
    since they share the same phase factor. There may also be an intercavity
    coupling $t_\text{ph}$ mediated by coupling of the QDs via the SC leads.  In
    the absence of magnetic fields, $t_\text{ph}$, too, is real. We note that
    Eq.~\eqref{eq-hphoton} is meaningful only if $|\omega\pm t_\text{ph}|>
    |\Delta_\text{ph}\pm\Lambda_\text{ph}|$. This is fulfilled in the
    offresonant regime, where $|\omega|$ is large compared to the other
    parameters in Eq.~\eqref{eq-hphoton}.

    A generalized Bogoliubov transformation to the new bosonic fields 
    $\pi_{\mu\nu}=
            [u_\nu(a^\dagger_{1\mu}+\nu a^\dagger_{2\mu})
            +v_\nu(\nu a_{1\bar\mu}+a_{2\bar\mu})]
            /(v_\nu^2-|u_\nu|^2)^{1/2}$
    with $u_\pm=\Delta_\text{ph}\pm\Lambda_\text{ph}$ and 
    $v_\pm=(\hbar\omega\pm t_\text{ph})+\sqrt{(\hbar\omega\pm t_\text{ph})^2
        -|\Delta_\text{ph}\pm\Lambda_\text{ph}|^2}$
    diagonalizes the Hamiltonian~\eqref{eq-hphoton}, such that
    \begin{equation}
        H_\text{ph}=\sum_{\mu\nu}E_\nu\pi^\dagger_{\mu\nu}\pi_{\mu\nu},
    \end{equation}
    where $E_\pm=\sqrt{(\hbar\omega\pm t_\text{ph})^2-
    |\Delta_\text{ph}\pm\Lambda_\text{ph}|^2}$. The ground state $\Ket{G}$ is
    fixed by the condition $\pi_{\mu\nu}\Ket{G}=0$ for all $\mu,\nu$.  In terms
    of the original photons this implies
    \begin{align}
        \Ket{G}=\frac{1}{v_+ v_-}&\exp\Bigl[
            -\frac{1}{2}\Bigl(\frac{u_+}{v_+}-\frac{u_-}{v_-}\Bigr)
                (a^\dagger_{1R}a^\dagger_{1L}+a^\dagger_{2R}a^\dagger_{2L})
            \notag\\
            -&\frac{1}{2}\Bigl(\frac{u_+}{v_+}+\frac{u_-}{v_-}\Bigr)
                (a^\dagger_{1R}a^\dagger_{2L}+a^\dagger_{1L}a^\dagger_{2R})
        \Bigr]\Ket{},
        \label{eq-groundstate}
    \end{align}
    where $\Ket{}$ denotes the photonic vacuum. The ground state
    \eqref{eq-groundstate} is a squeezed state \cite{vogel01} of local and
    nonlocal entangled photon pairs. At low temperatures and with a sufficiently
    high cavity quality factor, $\Ket{G}$ is the dominant contribution to the
    photonic density matrix.

    \emph{Photon detection and Bell test.}---%
    We use a positive operator valued measure to model the photon detection. The
    probability to detect at least one photon in cavity $i$ with polarization
    $\xi$ if $n_{i\xi}$ photons are present, is given by
    $P^+_{i\xi}=1-(1-\gamma)^{n_{i\xi}}$, where $\gamma\in[0,1]$ is the
    detection efficiency. With probability
    $P^-_{i\xi}=1-P^+_{i\xi}$, no photon is detected \cite{vivoli14}. More
    generally, a photon with the arbitrary polarization $\alpha$ is detected
    with probability $P^+_{i\alpha}=1-(1-\gamma)^{\tilde a^\dagger_{iR}\tilde
    a_{iR}}$, where $\tilde a_{1R}=a_{1R}\cos\alpha-a_{1L}\sin\alpha$ and
    $\tilde a_{1L}=a_{1R}\sin\alpha+a_{1L}\cos\alpha$.

    \begin{figure*}
        \includegraphics{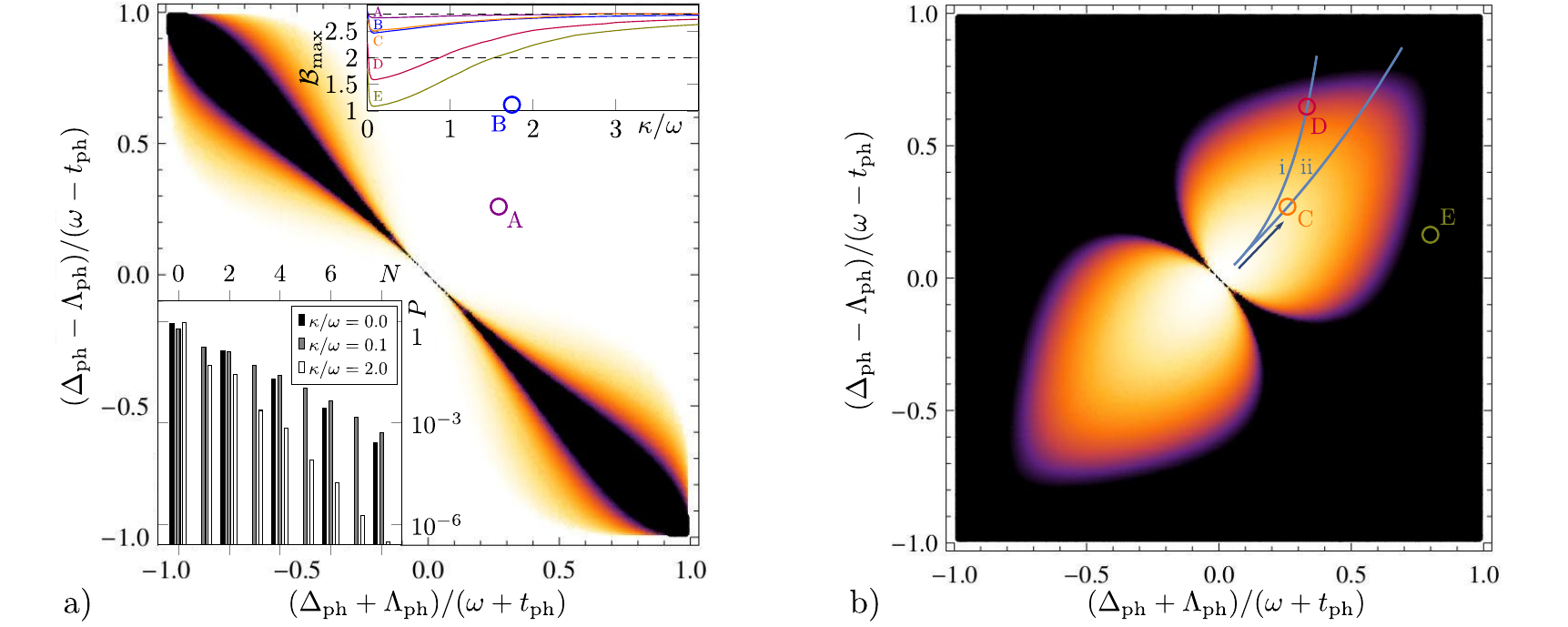}
        \caption{Maximum violation of the CHSH Bell inequality depending 
            on the local and nonlocal pair creation amplitudes
            $\Lambda_\text{ph}$ and $\Delta_\text{ph}$. In black regions
            entanglement cannot be detected, $\mathcal{B}_\text{max}\le2$. White
            regions correspond to $\mathcal{B}_\text{max}\ge2\sqrt{2}$. a)
            Perfect detectors, $\gamma=1$, isolate the contribution of a single
            nonlocal pair and entanglement is detectable for almost all
            parameters. Right inset:  cavity losses affect
            $\mathcal{B}_\text{max}$ (shown for different points A-E in the
            parameters space), possibly reducing it below the critical value 2
            (lower dashed line). At high losses, however, all surviving
            coincidences are due to nonlocal pairs and $\mathcal{B}_\text{max}$
            approaches the two-particle maximum value $2\sqrt{2}$ (upper dashed
            line). Left inset: probabilities $P$ to find $N$ photons in the
            cavities at point B/D in the parameter space for different loss
            rates $\kappa$. At finite cavity losses, odd number states are
            populated, because photon pairs are broken up. b) Lossy detectors,
            $\gamma=10^{-4}$ (see text), are able to detect entanglement if the
            admixture of local pairs is sufficiently small. The lines indicate
            the parameters obtained from the microscopic model if ECT is (i) as
            strong as CAR or (ii) half as strong as CAR. Following these lines
            in the direction of the arrow corresponds to different cavity
            detunings, $V_\text{sd}=\hbar\omega_0/e+\delta V_\text{sd}$, where
            $\delta V_\text{sd}$ increases from $-1~\mu$eV to $1~\mu$eV.}
        \label{fig-bell}
    \end{figure*}

    As a measure of entanglement we employ a CHSH-type Bell inequality.  We
    define four observables $A_1$, $B_1$, $C_2$ and $D_2$ with measurement
    outcomes $\pm1$, where the subscript indicates in which cavity the operator
    acts. The quantity $\mathcal{B}=A_1 C_2+A_1 D_2+B_1 C_2-B_1 D_2=
    A_1(C_2+D_2)+B_1(C_2-D_2)$ can take the values $\pm2$ and hence the
    expectation value $\langle\mathcal{B}\rangle$ lies within $[-2,2]$, provided
    that the observables $A_1$ to $D_2$ have definite values independent of any
    measurement \cite{clauser69}. We adopt the following scheme: $A_1\equiv
    A^\alpha_1$ is $(-)1$ if only photons with polarization $\alpha(+\pi/2)$ are
    detected in cavity 1. $B_1$ is the same measurement rotated by $45^\circ$,
    i.e., $B_1\equiv B_1^\alpha=A_1^{\alpha+\pi/4}$. Likewise $C_2^\beta$ and
    $D_2^\beta$ measure the photons in cavity 2. Inconclusive events, in which 
    no photons, or photons of both polarizations are detected on one side, are
    discarded. With the nonlocal correlations
    $\mathcal{C}(\alpha,\beta)=\sum_{\mu\nu}\mu\nu
    P^{\mu\nu}_{\alpha\beta}/\sum_{\mu\nu}P^{\mu\nu}_{\alpha\beta}$ the Bell
    parameter can be expressed as
    \begin{equation}
        \mathcal{B}_\alpha=\mathcal{C}(0,\alpha)
            +\mathcal{C}(0,\alpha+\frac{\pi}{4})
            +\mathcal{C}(\frac{\pi}{4},\alpha)
            -\mathcal{C}(\frac{\pi}{4},\alpha+\frac{\pi}{4}).
    \end{equation}
    In this notation, $P_{\alpha\beta}^{\mu\nu}$ is the probability to obtain
    the measurement result $\mu$ at cavity 1 when polarization detector 1 is set
    to the angle $\alpha$ and to obtain simultaneously the measurement result
    $\nu$ at cavity 2 when polarization detector 2 is set to the angle
    $\beta$. In an experiment, the probabilities are directly accessible by
    counting the respective coincidence events.  Formally, they are
    $P^{\mu\nu}_{\alpha\beta} =\Bra{G}P^\mu_{1\alpha}P^{\bar\mu}_{1\alpha+\pi/2}
    P^\nu_{2\beta}P^{\bar\nu}_{2\beta+\pi/2}\Ket{G}$. Owing to the structure of
    the ground state, the probabilities can be calculated
    to arbitrary precision, cf.~Supplemental Material \cite{suppl}, by finding
    the eigenvalues of a $4\times4$ matrix \cite{vivoli14}.

    For a given set of parameters $\omega$, $t_\text{ph}$, $\Delta_\text{ph}$ 
    and $\Lambda_\text{ph}$, the Bell test is conclusive only if there are
    angles $\alpha$, at which $|\mathcal{B}_\alpha|>2$. By inspection we find
    that, like in the two-photon case, $\mathcal{B}$ always takes its maximum
    for $\alpha=\frac{3\pi}{8}$.  The maximum value
    $\mathcal{B}_\text{max}:=\text{max}_\alpha|\mathcal{B_\alpha}|$ depends,
    however, on the detection efficiency $\gamma$.

    The most effective limit, $\gamma=1$, is hypothetical but instructive. It 
    corresponds to the situation in which all photons in the cavities can be
    measured with certainty, i.e., the cavities would have to be completely
    open. The ground state $\Ket{G}$ is a superposition of any number of local
    and of nonlocal photon pairs. With perfect detection, the vacuum state can
    be uniquely identified (no detector clicks) and is discarded according to
    the measurement scheme. Any state which contains at least one local pair
    has photons of both polarizations on one side, and is discarded, too. The
    remaining states contain one nonlocal pair, or, with an exponentially small
    amplitude, several nonlocal pairs.  Hence $\mathcal{B}_\text{max}$ is larger
    than the classical boundary 2 for almost any choice of parameters
    (Fig.~\ref{fig-bell}a). In a realistic situation, we model the finite cavity
    resonance width by choosing $\gamma<1$: even if the photodetectors
    themselves are flawless, during any coincidence interval $\Delta t$ the
    probability for a single photon to leave its cavity is $1-e^{-\kappa\Delta
    t}\gtrsim\gamma$, where we can express the loss rate
    $\kappa=(\omega+eV_\text{sd})/Q$ through the quality factor $Q$. Thus
    $\gamma$ can safely be assumed to be larger than
    $10^{-4}$ \footnote{For this estimate, we assume a worst-case set of 
        parameters: a high quality factor $Q=10^6$, infrared light $f=10$~THz
        and a short coincidence interval $\Delta t=1$~ps.}. 
    In the Bell measurement, only one photon of a {\it local pair} might be
    detected and believed to belong to a {\it nonlocal pair}. At too large
    values of $\Lambda_\text{ph}$, $\mathcal{B}_\text{max}$ is thus reduced
    below 2 and the entanglement of the nonlocal pairs -- although present --
    cannot be detected anymore (Fig.~\ref{fig-bell}b). Importantly, irrespective
    of $\gamma$, the Bell inequality is violated only if the nonlocal pairing
    amplitude $\Delta_\text{ph}$ is nonzero.

    \emph{Cavity leakage.}---%
    Besides affecting the detection efficiency, cavity losses also modify the
    steady state. This influences the Bell measurement in two ways: entangled
    pairs can be broken up leaving behind unpaired photons. If their detection
    coincides with other photons, $\mathcal{B}_\text{max}$ is reduced. On the
    other hand, $\mathcal{B}_\text{max}$ is increased because the probability to
    have additional local pairs in the cavities is reduced. We model cavity
    losses with a Lindblad master equation \cite{gardiner00} and solve it
    numerically \cite{suppl} for the steady state. When the loss rate $\kappa$
    dominates over the induced pairing amplitudes
    $|\Delta_\text{ph}|,|\Lambda_\text{ph}|$ (the most likely situation), in
    total $\mathcal{B}_\text{max}$ is increased (Fig.~\ref{fig-bell}a, inset).

    To summarize, the Bell inequality is violated robustly over a broad range of
    configurations, but only if the nonlocal pairing amplitude
    $\Delta_\text{ph}$ is nonzero. Since $\Delta_\text{ph}$ is directly related
    to CAR (see below), observing $\mathcal{B}_\text{max}>2$ is an experimental
    proof of coherent CP splitting.

    \emph{Microscopic model.}---%
    In the remainder of the text, we derive the effective photonic Hamiltonian
    \eqref{eq-hphoton} microscopically. For the n-side, consider a double
    QD coupled to a SC lead, where the SC gap $\Delta=|\Delta|e^{i\phi}$ is the
    largest energy, $|\Delta|\gg|\epsilon_i|,k_B T$, with $\epsilon_i$ the
    spin-degenerate levels of the two QDs $i=1,2$, and $T$ the temperature. We
    derive a Hamiltonian for the two QDs \cite{suppl} to second order in the
    electron tunneling between the QDs and the lead, 
    \begin{equation}
        H_e=\sum_{i\sigma}\epsilon_i d^\dagger_{i\sigma}d_{i\sigma}
            +\sum_\sigma\Bigl[
            \tilde\Delta \sigma d_{1\sigma}^\dagger d_{2\bar\sigma}^\dagger
            +\tilde t d^\dagger_{1\sigma}d_{2\sigma}+\text{h.c.}\Bigr],
        \label{eq-heff}
    \end{equation}
    where $d_{i\sigma}$ annihilates electrons with spin 
    $\sigma=\uparrow,\downarrow$ on QD $i$, and
    \begin{align}
        \label{eq-delta}
        \tilde\Delta&=-\frac{\Gamma}{2}
        \frac{e^{-\frac{r}{\pi\xi}+i\phi}}{k_F r}
        \Bigl[\sin(k_F r)+\frac{1-\cos(k_F r)}
                {\pi k_F\xi}\Bigr] \\
        \label{eq-lambda}
        \tilde t&=-\frac{\Gamma}{2}
        \frac{e^{-\frac{r}{\pi\xi}}}{k_F r}
            \Bigl[\frac{\sin(k_F r)}{\pi k_F\xi}-
                (1-\cos(k_F r))\Bigr],
    \end{align}
    where $r$ is the distance between the tunneling points to QD 1 and QD 2 from
    the lead, $\xi=\hbar v_F/(\pi|\Delta|)$ is the lead coherence length and
    $\Gamma=2\pi N(\epsilon_F)t_1 t_2$ is the normal state tunneling rate with
    $t_i$ the amplitude for an electron to tunnel from the SC lead to QD $i$ and
    $N$ the normal density of states in the SC lead.  We assume that double
    occupancy of a QD is forbidden by a large charging energy $U_i$
    \footnote{At finite Coulomb repulsion $U<|\Delta|$, the electron-hole system
        can be diagonalized perturbatively in $1/U$ or numerically. This
        substantially complicates the notation while not producing any new
        effects, so we focus on the (experimentally more relevant) case
        $U\gg|\Delta|$, in which ECT between the QDs via the SC lead is the
        dominant source of local photon pairs.}.
    On the p-side an equivalent Hamiltonian with $d_{i\sigma}\rightarrow 
    h_{i\sigma}=:d^{\text{(HH)}\dagger}_{i\bar\sigma}$ holds for the heavy holes
    (HH).  Here, $d^{\text{(HH)}\dagger}_{i\sigma}$ creates an electron with
    spin $\sigma$ on QD $i$ in the HH-band.
    
    The parameters $\tilde\Delta$ and $\tilde t$ capture the influence of the SC
    lead: $\tilde\Delta$ describes CAR, whereas $\tilde
    t$ describes ECT between the QDs.  Additionally, the onsite energies
    $\epsilon_i$ are renormalized.  Optimal splitting requires
    $|\tilde\Delta|\gg |\tilde t|$, because ECT allows for sequential emission
    into the same cavity. This is realized in the limit of a small Fermi wave
    vector $k_F r\rightarrow 0$, which is achievable in principle, since $k_F$
    describes the semiconductor in which $\Delta$ is proximity-induced.  For a
    more conservative estimate, if the separation of the tunnel contacts is
    large compared to the Fermi wavelength (but still smaller than $\xi$), it is
    evident that at best $|\tilde\Delta|\sim|\tilde t|$ can be realized. To be
    specific, we assume a large QD separation $k_F r\sim10$, $r/(\pi\xi)\sim0.5$
    and $\Gamma=500~\mu$eV ($\Gamma$ is limited by the gap in the SC, e.g.,
    1~meV for Nb). Then $\tilde\Delta$ and $\tilde t$ are on the order of
    $15~\mu$eV. We note that because of the oscillating contributions in
    Eqs.~\eqref{eq-delta} and~\eqref{eq-lambda} this may be fine-tuned via $k_F$
    in the SC lead.
    
    The cavities are coupled to the electronic system by electron-hole
    recombination. In dipole approximation with the usual selection rules for HH
    using the rotating wave approximation,
    \begin{equation}
        H_I=\sum_{i\sigma}g_{i\sigma}e^{-ieV_\text{sd}t/\hbar}
                d_{i\bar\sigma}h_{i\sigma}
                a^\dagger_{i\sigma} + \text{h.c.},
        \label{eq-hphotint}
    \end{equation}
    with the annihilation operators for photons $a_{i\xi}$ emitted along the
    quantization axis of angular momentum and with radiative couplings
    $g_{i\xi}$ for cavity $i$ and polarization $\xi$. For brevity we identify
    $\sigma=\;\uparrow$ with $\xi=R$ and $\sigma=\;\downarrow$ with $\xi=L$. The
    exponential accounts for the difference between the chemical potentials of
    n- and p-side set by the applied voltage bias. When it is gauged into the
    photon field $a\rightarrow e^{ieV_\text{sd}t/\hbar}a$, the complete photonic
    part of the Hamiltonian becomes
    \begin{equation}
        H_P=\sum_{i\xi}
            (\omega_0-eV_\text{sd}) a^\dagger_{i\xi} a_{i\xi}
            +\sum_{i\sigma}g_{i\sigma}
            d_{i\bar\sigma}h_{i\sigma}
            a^\dagger_{i\sigma} + \text{h.c.},
        \label{eq-hphot2}
    \end{equation}
    with the cavity resonance frequency $\omega_0$.

    \emph{Photoemission.}---%
    The ground state of the electronic system is a superposition of the empty
    state and the singlet state, $\Ket{0}_e=c^0_{e,0}\Ket{}
    +c^0_{e,s}\Ket{S}$ with $\Ket{}$ the empty dot, and $\Ket{S}:=2^{-1/2}(
    d^\dagger_{1\uparrow}d^\dagger_{2\downarrow}-
    d^\dagger_{1\downarrow}d^\dagger_{2\uparrow})\Ket{}$ the nonlocal
    singlet. The ground state energy is
    $E^0_e=\epsilon-\sqrt{\epsilon^2+2|\tilde\Delta|^2}$, where
    $\epsilon_{1,2}=:\epsilon\pm\delta$. The admixture of the singlet is controlled
    by the CAR amplitude $\tilde\Delta$, i.e., $|c^0_{e,s}|\rightarrow
    |c^0_{e,0}|$ as $|\tilde\Delta|/\epsilon\rightarrow\infty$. There are four
    relevant excited states,
    $\Ket{\sigma\pm}_e=c^\pm_{e,1}\Ket{1\sigma}+c^\pm_{e,2}\Ket{2\sigma}$ with
    energies $E^\pm_e=\epsilon\pm\sqrt{\delta^2+\tilde t^2}$, containing a
    single electron, which is delocalized due to ECT. Here,
    $\Ket{i\sigma}:=d^\dagger_{i\sigma}\Ket{}$. The same applies to the hole
    states (Fig.~\ref{fig-setup}). By electron-hole recombination and
    simultaneous emission of a photon as described by $H_I$, the n- and the
    p-side can both be excited from the ground state $\Ket{0}_e\Ket{0}_h$ into
    an excited state $\Ket{\sigma\mu}_e\Ket{\bar\sigma\nu}_h$, where
    $\mu,\nu=\pm$. Parity being protected, in order to go back to the ground
    state, this has to be followed by either a reabsorption, or by the emission
    of a second photon.  This is possible because of the induced superconducting
    pairing, which breaks particle number conservation. If the excitation
    energies $\Delta E^{\mu\nu}=E^\mu_e+E^\nu_h-E^0_e-E^0_h\gtrsim
    \epsilon_e+\epsilon_h$, are larger than the cavity linewidth, this is a
    virtual process and photons are always emitted in pairs as desired.
    Importantly, the electron-hole system undergoes a closed cycle, so the
    entanglement is completely transferred \cite{titov05} onto the photons 
    \footnote{Because of parity conservation we expect the photons to be 
        entangled even if the cavity linewidth $\kappa$ is larger than 
        $\Delta E^{\mu\nu}$ (but smaller than $\Delta$ in the SC leads and $U$). 
        This, however, leads to a different regime in which single photons can
        be emitted sequentially, and requires a separate calculation.}.
    To be specific, at $\tilde\Delta=15~\mu$eV, a gate voltage needs to be
    applied to the QDs such that (i) $\epsilon=30~\mu$eV if 
    $\tilde t=\tilde\Delta$, or such that (ii) $\epsilon=20~\mu$eV if $\tilde
    t=\tilde\Delta/2$. Here we assume $\delta=0$, and electron-hole symmetry for
    simplicity, and use a linewidth on the order of $10$~GHz, achievable for all
    common nanocavity flavors \cite{hennessy07,srinivasan07,loo10,ohta11}. In
    principle, $\Delta E^{\mu\nu}$ can be made arbitrarily large (within the
    lead SC gap) to account for higher cavity losses by increasing $\epsilon$.
    Since this, however, also reduces the CAR amplitude, it comes at the expense
    of a lower pair emission rate. With stronger CAR,
    $\tilde\Delta\gtrsim30~\mu$eV, the QDs can even be brought into resonance
    with the leads, $\epsilon=0$, to optimize the emission rate.

    Quantitatively, the emission cycle is conveniently described with a standard
    Schrieffer-Wolff transformation \cite{schrieffer66}, yielding the effective
    photonic model \eqref{eq-hphoton}, with the pair emission amplitudes
    $\Delta_\text{ph}$, $\Lambda_\text{ph}\sim g(g/\Delta E)(\tilde\Delta/
    \epsilon)^2$. This corresponds to a total emission rate on the order of
    $10^{-3}g/\hbar$ if $\kappa>|\Delta_\text{ph}|,|\Lambda_\text{ph}|$. The
    full expressions are given in the Supplemental Material \cite{suppl}. As
    expected, intercavity hopping $t_\text{ph}$ and local pair injection
    $\Lambda_\text{ph}$ vanish for $\tilde t=0$. More importantly, even at
    finite $\tilde t$, we obtain effective parameters well within the region
    $\mathcal{B}_\text{max}>2$. They are shown for the QD parameters (i) and
    (ii) at a QD-cavity coupling strength $g/(2\pi \hbar)=3~$GHz
    \cite{srinivasan07,loo10}, in Fig.~\ref{fig-bell}b.

    In conclusion, we have investigated a coherent and continuous source of
    nonlocal photon pairs emitted from two optical QDs into two nanocavities
    embedded in a superconducting p-n junction. We showed that a Bell
    measurement can be used to investigate whether the light produced is
    entangled.  Detection of these entangled photons would be a proof for spin
    coherence of Cooper pairs split over two QDs. The device does not require a
    driving scheme but only dc-voltages.

    \begin{acknowledgments}
        We thank J.\ C.\ Budich and F.\ Hassler for helpful discussions and
        acknowledge support from the EU-FP7 project SE2ND, No. 271554 and the
        DFG grant No. RE 2978/1-1.
    \end{acknowledgments}

    \setcounter{figure}{0}
    \renewcommand{\thefigure}{S\arabic{figure}}
    \setcounter{equation}{0}
    \renewcommand{\theequation}{S\arabic{equation}}
    \section{Supplemental material}
        \subsection{Calculation of the detection probabilities}
            The detection probabilities $P^{\mu\nu}_{\alpha\beta}$ decompose
            into sums of expressions of the form
            \begin{align}
                P_\eta&=\Bra{G}
                    \eta_{1R}^{\tilde a^\dagger_{1R}\tilde a_{1R}}
                    \eta_{1L}^{\tilde a^\dagger_{1L}\tilde a_{1L}}
                    \eta_{2R}^{\tilde a^\dagger_{2R}\tilde a_{2R}}
                    \eta_{2L}^{\tilde a^\dagger_{2L}\tilde a_{2L}}
                    \Ket{G} \notag\\
                    &=\Bigl|
                    \eta_{1R}^{\frac{1}{2}\tilde a^\dagger_{1R}\tilde a_{1R}}
                    \eta_{1L}^{\frac{1}{2}\tilde a^\dagger_{1L}\tilde a_{1L}}
                    \eta_{2R}^{\frac{1}{2}\tilde a^\dagger_{2R}\tilde a_{2R}}
                    \eta_{2L}^{\frac{1}{2}\tilde a^\dagger_{2L}\tilde a_{2L}} 
                    \notag\\
                    &\hspace{1cm}
                    \exp\Bigl[A(a_{1R}^\dagger a_{1L}^\dagger+
                    a_{2R}^\dagger a_{2L}^\dagger) \notag\\
                    &\hspace{1.5cm}+
                    B(a_{1R}^\dagger a_{2L}^\dagger+
                    a_{1L}^\dagger a_{2R}^\dagger)\Bigr]\Ket{}
                    \Bigr|^2,
            \end{align}
            where $\eta_{i\xi}=1-\gamma$ or $\eta_{i\xi}=1$. Employing the
            unitary transformation
            \begin{equation*}
                \begin{pmatrix} \tilde a_{1R} \\ \tilde a_{1L} \\ 
                    \tilde a_{2R} \\ \tilde a_{2L} 
                \end{pmatrix}
                =U\begin{pmatrix} 
                    a_{1R} \\ a_{1L} \\ a_{2R} \\ a_{2L} 
                \end{pmatrix},
            \end{equation*}
            \begin{equation}
                \small
                U=\begin{pmatrix} \cos\alpha & -e^{i\phi}\sin\alpha & 0 & 0 \\ 
                    e^{-i\phi}\sin\alpha & \cos\alpha & 0 & 0 \\
                    0 & 0 & \cos\beta & -e^{i\theta}\sin\beta \\ 
                    0 & 0 & e^{-i\theta}\sin\beta & \cos\beta \\
                \end{pmatrix},
            \end{equation}
            together with the operator identity $x^{a^\dagger a}f(a^\dagger)=
            f(xa^\dagger)x^{a^\dagger a}$ \cite{sekatski10}, and using that
            $x^{a^\dagger a}\Ket{}=\Ket{}$, we rewrite
            \begin{align}
                P_\eta&\propto\Bigl|\exp[\frac{1}{2}
                    \begin{pmatrix} \tilde a^\dagger_{1R} & 
                        \tilde a^\dagger_{1L} & \tilde a^\dagger_{2R} & 
                        \tilde a^\dagger_{2L} 
                    \end{pmatrix}
                    D^\text{T}U^\text{T}MUD
                    \begin{pmatrix} \tilde a^\dagger_{1R} \\ 
                        \tilde a^\dagger_{1L} \\ \tilde a^\dagger_{2R} \\ 
                        \tilde a^\dagger_{2L} 
                    \end{pmatrix}]\Ket{}\Bigr|^2 \notag\\
                &=\Bigl|\exp[\frac{1}{2}\sum_{i_1}^4\lambda_i b^\dagger_i b_i]
                    \Ket{}\Bigr|^2,
            \end{align}
            where
            \begin{equation}
                D=\begin{pmatrix}
                    \sqrt{\eta_{1R}} & 0 & 0 & 0 \\
                    0 & \sqrt{\eta_{1L}} & 0 & 0 \\
                    0 & 0 & \sqrt{\eta_{2R}} & 0 \\
                    0 & 0 & 0 & \sqrt{\eta_{2L}} 
                \end{pmatrix},
            \end{equation}
            and
            \begin{equation}
                M=\begin{pmatrix}
                    0 & A & 0 & B \\ A & 0 & B & 0 \\
                    0 & A & 0 & B \\ A & 0 & B & 0
                \end{pmatrix},
            \end{equation}
            and $\lambda_i$ are the eigenvalues of $D^\text{T}U^\text{T}MUD$.
            The new bosons $b_i$ are independent, so
            \begin{equation}
                P_\eta\propto\prod_{i=1}^4\Bra{}
                    e^{\frac{1}{2}\lambda^*_i b_i b_i}
                    e^{\frac{1}{2}\lambda b^\dagger_i b^\dagger_i}
                \Ket{}
                =\prod_{i=1}^4\Bigl(1-|\lambda_i|^2\Bigr).
                \label{eq-probfinal}
            \end{equation}
            The last equality can be found by inserting the definition of the
            exponential as a series. If $A$ and $B$ have the same phase, it
            drops out in Eq.~\eqref{eq-probfinal}.

        \subsection{Lindblad master equation}
            To study the influence of cavity losses, we solve the Lindblad
            master equation \cite{gardiner00}
            \begin{equation}
                \dot\rho=\mathcal{L}\rho=-i[H,\rho]+\sum_{i\xi}\kappa
                \left(2a_{i\xi}\rho a_{i\xi}^\dagger
                    -a_{i\xi}^\dagger a_{i\xi}\rho
                    -\rho a_{i\xi}^\dagger a_{i\xi}\right)
            \end{equation}
            numerically for the steady state, $\dot\rho_\text{steady}=0$. Here,
            $\rho$ is the density matrix and $\mathcal{L}$ is the Liouvillian
            superoperator, defined by the rightmost equality sign. The system
            Hamiltonian $H$ is given by Eq.~\eqref{eq-hphoton}. The
            infinite-dimensional bosonic Hilbert space has to be truncated. We
            choose a cutoff at $N_\text{max}=4$ bosons per mode $(i\xi)$. The
            cutoff independence can be checked by comparing to a higher cutoff
            $N_\text{max}=5$, comparing the solutions in a photon basis to the
            solution in a Bogoliubov basis and by comparing the
            $\kappa\rightarrow0$ result with the exact calculation. In order to
            obtain the steady state we either explicitly propagate an arbitrary
            initial state in time until convergence or algebraically solve for
            the eigenvector of $\mathcal{L}$ which has the smallest eigenvalue
            (always numerical~0).

        \subsection{Proximity effect}
            In this section we derive the pairing amplitudes $\tilde\Delta$ and
            $\tilde t$ on the QDs, Eqs.~\eqref{eq-delta} and~\eqref{eq-lambda}.
            This is done most conveniently by performing a Schrieffer-Wolff
            transformation on an operator level via projection operators,
            following Ref.~\cite{recher10}. It applies independently to n- and
            p-side.

            We consider the full device depicted in Fig.~\ref{fig-setup}. Two
            optical QDs are coupled to a SC lead. 
            Specifically,
            \begin{align}
                H&=H_0+H_T \\
                H_0&=\sum_{{\bf k}\sigma}E_{\bf k}
                    \gamma^\dagger_{{\bf k}\sigma}\gamma_{{\bf k}\sigma}
                    +\sum_{i\sigma}\epsilon_i
                        d^\dagger_{i\sigma}d_{i\sigma} \notag\\
                    &+\sum_i U_i
                        d^\dagger_{i\uparrow}d^\dagger_{i\downarrow}
                        d_{i\downarrow}d_{i\uparrow} \\
                H_T&=\sum_{i\sigma} t_i
                    d^\dagger_{i\sigma}\psi_\sigma({\bf x}_i)
                        +\text{h.c.},
            \end{align}
            where $E_{\bf k}=\sqrt{(\epsilon_k-\mu)^2+|\Delta|^2}$ is the 
            dispersion of quasiparticles in the lead with gap energy $|\Delta|$
            annihilated by $\gamma$; $\epsilon_i$ and $U_i\rightarrow\infty$ are
            the onsite energy measured from the chemical potential and the local
            Coulomb repulsion of electrons with spin
            $\sigma\in\{\uparrow,\downarrow\}$ on QD $i\in\{1,2\}$ annihilated
            by $d$. We assume that there is only a single relevant level on each
            dot. Tunneling from point ${\bf x}_i$ in the lead onto QD $i$ has
            the amplitude $t_i\in\mathbb{R}$ (no magnetic fields).  The
            operators $\psi$ and $\gamma$ are related through the Bogoliubov
            transformation $\psi_\sigma({\bf x})=\sum_{\bf k}e^{i{\bf
            kx}}(u^*_{\bf k} \gamma_{{\bf k}\sigma}+\sigma v_{\bf
            k}\gamma^\dagger_{-{\bf k}\bar\sigma})$, where the coherence
            factors obey $|v_{\bf k}|^2=1-|u_{\bf k}|^2
            =(1-(\epsilon_{\bf k}-\mu)/E_{\bf k})/2$ and
            $v^*_{\bf k}u_{\bf k}=-|v_{\bf k}||u_{\bf k}|e^{-i\phi}$ with the
            SC phase $\phi=\arg\Delta$.

            \begin{figure}
                \includegraphics{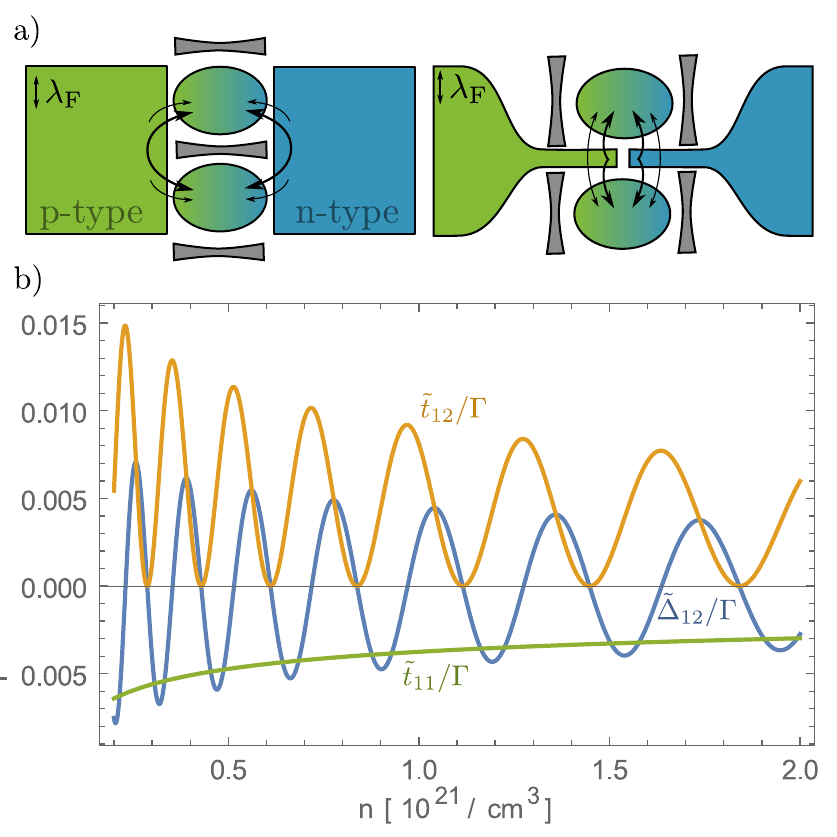}
                \caption{a) Two different device geometries. Left panel: The
                    tunnel contacts between the SC leads and QDs are separated
                    by several Fermi wavelengths. ECT between the QDs is as
                    strong as CP splitting. Right panel: Electrons can tunnel
                    into both QDs from the same point in the SC leads. CP
                    splitting dominates over ECT.  b) Proximity induced
                    amplitudes of CAR, $\tilde\Delta_{12}$, and ECT,
                    $\tilde t_{12}$, and the renormalization of the on-site 
                    energy $\tilde t_{11}$ at different carrier concentrations 
                    $n$.  When optimizing the CP splitting amplitude, ECT cannot
                    be neglected. The QDs are separated by $r=100$~nm and the SC
                    coherence length is $\xi=200$~nm (arbitrarily chosen).}
                \label{fig-layout}
            \end{figure}

            Defining a set of projection operators,
            \begin{equation}
                P=\prod_{{\bf k}\sigma}\Bigl(
                    1-\gamma^\dagger_{{\bf k}\sigma}\gamma_{{\bf k}\sigma}\Bigr)
                \hspace{1cm}
                Q=1-P,
            \end{equation}
            where $P$ projects onto the lead ground state, the Hamiltonian can
            be written as
            \begin{equation}
                H_\text{eff}=PHP+PHQ(E-QHQ)^{-1}QHP
                \label{eq-hprojected}
            \end{equation}
            in the low-energy subspace \cite{essler05}. Up to leading order in
            the tunneling amplitude $t_i$, this becomes
            \begin{equation}
                H_\text{eff}=H_0+\sum_{ij\sigma}\Bigl[
                    \tilde\Delta_{ij}
                    \sigma d_{i\sigma}^\dagger d_{j\bar\sigma}^\dagger
                        +\text{h.c.}
                +\tilde t_{ij} d^\dagger_{i\sigma}d_{j\sigma}\Bigr]
            \end{equation}
            with the (C)AR amplitude
            \begin{equation}
                \tilde\Delta_{ij}=\sum_{\bf k}
                    \frac{t_i t_j}{E_{\bf k}}
                    \psi_{\bf k}({\bf x}_i)\psi_{-{\bf k}}({\bf x}_j)
                    |u_{\bf k}||v_{\bf k}|e^{i\phi},
                \label{eq-deltainter}
            \end{equation}
            and the ECT amplitude
            \begin{equation}
                \tilde t_{ij}=-\sum_{\bf k}
                    \frac{t_i t_j}{E_{\bf k}}
                    \psi_{{\bf k}\sigma}({\bf x}_i)
                    \psi^*_{{\bf k}\sigma}({\bf x}_j)
                    (|u_{\bf k}|^2-|v_{\bf k}|^2),
                \label{eq-lambdainter}
            \end{equation}
            where we have used that $E_{\bf k}\sim|\Delta|\gg|\epsilon|$.
            Starting from Eq.~\eqref{eq-hprojected} higher-order terms can be
            included systematically, but they do not contain new effective
            processes, so it is safe to ignore them.  Linearizing the bare
            spectrum of the lead $\epsilon_{\bf k}$ within the SC gap, the
            momentum summations can be performed as integrals over the density
            of states and we obtain $\tilde\Delta\equiv\tilde\Delta_{12}$ and
            $\tilde t\equiv\tilde t_{12}$ as given in Eqs.~\eqref{eq-delta}
            and~\eqref{eq-lambda}. The renormalization of the onsite energies by
            $\tilde t_{ii}$ is obtained as the $r\rightarrow 0$ limit of
            Eq.~\eqref{eq-lambda}, but can be absorbed into $\epsilon$.  The
            dependency of the amplitudes on the carrier concentration is
            illustrated in Fig.~\ref{fig-layout}b.

        \subsection{Parameters of the effective model}
            Decoupling the low energy sector (the electron-hole system is in its
            ground state) from the high energy sector (the electron-hole system
            is in any other state) up to second order in $g_{i\xi}$, following, 
            e.g., Ref.~\cite{winkler03}, we find
            \begin{align}
                H_\text{ph}=
                    \sum_{i\xi}(\omega_0&-eV_\text{sd})
                        a^\dagger_{i\xi}a_{i\xi}
                        +\sum_{ij\xi}M_{ij\xi}
                        a^\dagger_{i\xi}a_{j\xi} \notag\\
                        &+\sum_{ij\xi}\frac{\tilde M_{ij\xi}}{2}
                            a_{i\xi}a_{i\bar\xi} 
                    +\text{h.c.},
                \label{eq-heff-s}
            \end{align}
            where
            \begin{widetext}
            \begin{align}
                M_{ij\xi}&=g_{i\xi}g^*_{j\xi}\sum_{\mu\nu=\pm}\Big(
                    \frac{|c^0_{e,0}|^2|c^0_{h,0}|^2
                            c^\mu_{e,i}c^{\mu*}_{e,j}
                            c^\nu_{h,i}c^{\nu*}_{h,j}}
                        {(\omega_0-eV_\text{sd})-\Delta E^{\mu\nu}}
                    -\frac{1}{4}\frac{|c^0_{e,s}|^2|c^0_{h,s}|^2
                            c^{\mu*}_{e,\bar i}c^\mu_{e,\bar j}
                            c^{\nu*}_{h,\bar i}c^\nu_{h,\bar j}}
                        {(\omega_0-eV_\text{sd})+\Delta E^{\mu\nu}}
                \Big) \\
                \tilde M_{ij\xi}&=g^*_{i\xi}g^*_{j\bar\xi}
                \sum_{\mu\nu=\pm}
                    c^{0*}_{e,s}c^0_{e,0}
                    c^{0*}_{h,s}c^0_{h,0}
                    c^{\nu*}_{h,j}c^\nu_{h,\bar i}
                    c^{\mu*}_{e,j}c^\mu_{e,\bar i}
                    \frac{\Delta E^{\mu\nu}}
                        {(\Delta E^{\mu\nu})^2-(\omega_0-eV_\text{sd})^2}.
            \end{align}
            \end{widetext}
            The amplitudes $c$ are obtained exactly by diagonalizing the
            $2\times 2$ blocks of the effective QD Hamiltonian~\eqref{eq-heff}.
            At symmetric gating, $\delta=0$, as discussed in the main text the
            excited state amplitudes $c_{e/h,i}^\pm$ are on the order of 1 and
            the ground state amplitudes fulfill $|c_{e/h,s}^{0*}c_{e/h,0}^0|=
            |\tilde\Delta|/(2|\epsilon|)
            +\mathcal{O}(|\tilde\Delta|^2/\epsilon^2)$.
            With two identical QDs, two identical cavities, and polarization
            independent radiative couplings $g$, Eq.~\eqref{eq-heff-s} is
            equivalent to the effective photonic model~\eqref{eq-hphoton},
            with $\omega=\omega_0-eV_\text{sd}+M_{i=j}$, $t_\text{ph}=M_{i\neq
            j}$, $\Lambda_\text{ph}=\tilde M_{i=j}$ and $\Delta_\text{ph}=\tilde
            M_{i\neq j}$.


\begin{thebibliography}{70}%
\makeatletter
\providecommand \@ifxundefined [1]{%
 \@ifx{#1\undefined}
}%
\providecommand \@ifnum [1]{%
 \ifnum #1\expandafter \@firstoftwo
 \else \expandafter \@secondoftwo
 \fi
}%
\providecommand \@ifx [1]{%
 \ifx #1\expandafter \@firstoftwo
 \else \expandafter \@secondoftwo
 \fi
}%
\providecommand \natexlab [1]{#1}%
\providecommand \enquote  [1]{``#1''}%
\providecommand \bibnamefont  [1]{#1}%
\providecommand \bibfnamefont [1]{#1}%
\providecommand \citenamefont [1]{#1}%
\providecommand \href@noop [0]{\@secondoftwo}%
\providecommand \href [0]{\begingroup \@sanitize@url \@href}%
\providecommand \@href[1]{\@@startlink{#1}\@@href}%
\providecommand \@@href[1]{\endgroup#1\@@endlink}%
\providecommand \@sanitize@url [0]{\catcode `\\12\catcode `\$12\catcode
  `\&12\catcode `\#12\catcode `\^12\catcode `\_12\catcode `\%12\relax}%
\providecommand \@@startlink[1]{}%
\providecommand \@@endlink[0]{}%
\providecommand \url  [0]{\begingroup\@sanitize@url \@url }%
\providecommand \@url [1]{\endgroup\@href {#1}{\urlprefix }}%
\providecommand \urlprefix  [0]{URL }%
\providecommand \Eprint [0]{\href }%
\providecommand \doibase [0]{http://dx.doi.org/}%
\providecommand \selectlanguage [0]{\@gobble}%
\providecommand \bibinfo  [0]{\@secondoftwo}%
\providecommand \bibfield  [0]{\@secondoftwo}%
\providecommand \translation [1]{[#1]}%
\providecommand \BibitemOpen [0]{}%
\providecommand \bibitemStop [0]{}%
\providecommand \bibitemNoStop [0]{.\EOS\space}%
\providecommand \EOS [0]{\spacefactor3000\relax}%
\providecommand \BibitemShut  [1]{\csname bibitem#1\endcsname}%
\let\auto@bib@innerbib\@empty
%</preamble>
\bibitem [{\citenamefont {Nielsen}\ and\ \citenamefont
  {Chuang}(2000)}]{nielsen00}%
  \BibitemOpen
  \bibfield  {author} {\bibinfo {author} {\bibfnamefont {M.~A.}\ \bibnamefont
  {Nielsen}}\ and\ \bibinfo {author} {\bibfnamefont {I.~L.}\ \bibnamefont
  {Chuang}},\ }\href@noop {} {\emph {\bibinfo {title} {Quantum Computation and
  Quantum Information}}}\ (\bibinfo  {publisher} {Cambridge University Press,
  Cambridge, England},\ \bibinfo {year} {2000})\BibitemShut {NoStop}%
\bibitem [{\citenamefont {Choi}\ \emph {et~al.}(2000)\citenamefont {Choi},
  \citenamefont {Bruder},\ and\ \citenamefont {Loss}}]{choi00}%
  \BibitemOpen
  \bibfield  {author} {\bibinfo {author} {\bibfnamefont {M.-S.}\ \bibnamefont
  {Choi}}, \bibinfo {author} {\bibfnamefont {C.}~\bibnamefont {Bruder}}, \ and\
  \bibinfo {author} {\bibfnamefont {D.}~\bibnamefont {Loss}},\ }\href {\doibase
  10.1103/PhysRevB.62.13569} {\bibfield  {journal} {\bibinfo  {journal} {Phys.
  Rev. B}\ }\textbf {\bibinfo {volume} {62}},\ \bibinfo {pages} {13569}
  (\bibinfo {year} {2000})}\BibitemShut {NoStop}%
\bibitem [{\citenamefont {Torr\`es}\ and\ \citenamefont
  {Martin}(1999)}]{torres99}%
  \BibitemOpen
  \bibfield  {author} {\bibinfo {author} {\bibfnamefont {J.}~\bibnamefont
  {Torr\`es}}\ and\ \bibinfo {author} {\bibfnamefont {T.}~\bibnamefont
  {Martin}},\ }\href {\doibase 10.1007/s100510051010} {\bibfield  {journal}
  {\bibinfo  {journal} {Eur. Phys. J. B}\ }\textbf {\bibinfo {volume} {12}},\
  \bibinfo {pages} {319} (\bibinfo {year} {1999})}\BibitemShut {NoStop}%
\bibitem [{\citenamefont {Falci}\ \emph {et~al.}(2001)\citenamefont {Falci},
  \citenamefont {Feinberg},\ and\ \citenamefont {Hekking}}]{falci01}%
  \BibitemOpen
  \bibfield  {author} {\bibinfo {author} {\bibfnamefont {G.}~\bibnamefont
  {Falci}}, \bibinfo {author} {\bibfnamefont {D.}~\bibnamefont {Feinberg}}, \
  and\ \bibinfo {author} {\bibfnamefont {F.~W.~J.}\ \bibnamefont {Hekking}},\
  }\href {\doibase 10.1209/epl/i2001-00303-0} {\bibfield  {journal} {\bibinfo
  {journal} {Europhys. Lett.}\ }\textbf {\bibinfo {volume} {54}},\ \bibinfo
  {pages} {255} (\bibinfo {year} {2001})}\BibitemShut {NoStop}%
\bibitem [{\citenamefont {Recher}\ \emph {et~al.}(2001)\citenamefont {Recher},
  \citenamefont {Sukhorukov},\ and\ \citenamefont {Loss}}]{recher01}%
  \BibitemOpen
  \bibfield  {author} {\bibinfo {author} {\bibfnamefont {P.}~\bibnamefont
  {Recher}}, \bibinfo {author} {\bibfnamefont {E.~V.}\ \bibnamefont
  {Sukhorukov}}, \ and\ \bibinfo {author} {\bibfnamefont {D.}~\bibnamefont
  {Loss}},\ }\href {\doibase 10.1103/PhysRevB.63.165314} {\bibfield  {journal}
  {\bibinfo  {journal} {Phys. Rev. B}\ }\textbf {\bibinfo {volume} {63}},\
  \bibinfo {pages} {165314} (\bibinfo {year} {2001})}\BibitemShut {NoStop}%
\bibitem [{\citenamefont {Lesovik}\ \emph {et~al.}(2001)\citenamefont
  {Lesovik}, \citenamefont {Martin},\ and\ \citenamefont
  {Blatter}}]{lesovik01}%
  \BibitemOpen
  \bibfield  {author} {\bibinfo {author} {\bibfnamefont {G.}~\bibnamefont
  {Lesovik}}, \bibinfo {author} {\bibfnamefont {T.}~\bibnamefont {Martin}}, \
  and\ \bibinfo {author} {\bibfnamefont {G.}~\bibnamefont {Blatter}},\ }\href
  {\doibase 10.1007/s10051-001-8675-4} {\bibfield  {journal} {\bibinfo
  {journal} {Eur. Phys. J. B}\ }\textbf {\bibinfo {volume} {24}},\ \bibinfo
  {pages} {287} (\bibinfo {year} {2001})}\BibitemShut {NoStop}%
\bibitem [{\citenamefont {Recher}\ and\ \citenamefont {Loss}(2002)}]{recher02}%
  \BibitemOpen
  \bibfield  {author} {\bibinfo {author} {\bibfnamefont {P.}~\bibnamefont
  {Recher}}\ and\ \bibinfo {author} {\bibfnamefont {D.}~\bibnamefont {Loss}},\
  }\href {\doibase 10.1103/PhysRevB.65.165327} {\bibfield  {journal} {\bibinfo
  {journal} {Phys. Rev. B}\ }\textbf {\bibinfo {volume} {65}},\ \bibinfo
  {pages} {165327} (\bibinfo {year} {2002})}\BibitemShut {NoStop}%
\bibitem [{\citenamefont {Bena}\ \emph {et~al.}(2002)\citenamefont {Bena},
  \citenamefont {Vishveshwara}, \citenamefont {Balents},\ and\ \citenamefont
  {Fisher}}]{bena02}%
  \BibitemOpen
  \bibfield  {author} {\bibinfo {author} {\bibfnamefont {C.}~\bibnamefont
  {Bena}}, \bibinfo {author} {\bibfnamefont {S.}~\bibnamefont {Vishveshwara}},
  \bibinfo {author} {\bibfnamefont {L.}~\bibnamefont {Balents}}, \ and\
  \bibinfo {author} {\bibfnamefont {M.~P.~A.}\ \bibnamefont {Fisher}},\ }\href
  {\doibase 10.1103/PhysRevLett.89.037901} {\bibfield  {journal} {\bibinfo
  {journal} {Phys. Rev. Lett.}\ }\textbf {\bibinfo {volume} {89}},\ \bibinfo
  {pages} {037901} (\bibinfo {year} {2002})}\BibitemShut {NoStop}%
\bibitem [{\citenamefont {Recher}\ and\ \citenamefont {Loss}(2003)}]{recher03}%
  \BibitemOpen
  \bibfield  {author} {\bibinfo {author} {\bibfnamefont {P.}~\bibnamefont
  {Recher}}\ and\ \bibinfo {author} {\bibfnamefont {D.}~\bibnamefont {Loss}},\
  }\href {\doibase 10.1103/PhysRevLett.91.267003} {\bibfield  {journal}
  {\bibinfo  {journal} {Phys. Rev. Lett.}\ }\textbf {\bibinfo {volume} {91}},\
  \bibinfo {pages} {267003} (\bibinfo {year} {2003})}\BibitemShut {NoStop}%
\bibitem [{\citenamefont {Levy~Yeyati}\ \emph {et~al.}(2007)\citenamefont
  {Levy~Yeyati}, \citenamefont {Bergeret}, \citenamefont {Martin-Rodero},\ and\
  \citenamefont {Klapwijk}}]{yeyati07}%
  \BibitemOpen
  \bibfield  {author} {\bibinfo {author} {\bibfnamefont {A.}~\bibnamefont
  {Levy~Yeyati}}, \bibinfo {author} {\bibfnamefont {F.~S.}\ \bibnamefont
  {Bergeret}}, \bibinfo {author} {\bibfnamefont {A.}~\bibnamefont
  {Martin-Rodero}}, \ and\ \bibinfo {author} {\bibfnamefont {T.~M.}\
  \bibnamefont {Klapwijk}},\ }\href {\doibase 10.1038/nphys621} {\bibfield
  {journal} {\bibinfo  {journal} {Nat. Phys.}\ }\textbf {\bibinfo {volume}
  {3}},\ \bibinfo {pages} {455} (\bibinfo {year} {2007})}\BibitemShut {NoStop}%
\bibitem [{\citenamefont {Cayssol}(2008)}]{cayssol08}%
  \BibitemOpen
  \bibfield  {author} {\bibinfo {author} {\bibfnamefont {J.}~\bibnamefont
  {Cayssol}},\ }\href {\doibase 10.1103/PhysRevLett.100.147001} {\bibfield
  {journal} {\bibinfo  {journal} {Phys. Rev. Lett.}\ }\textbf {\bibinfo
  {volume} {100}},\ \bibinfo {pages} {147001} (\bibinfo {year}
  {2008})}\BibitemShut {NoStop}%
\bibitem [{\citenamefont {Sato}\ \emph {et~al.}(2010)\citenamefont {Sato},
  \citenamefont {Loss},\ and\ \citenamefont {Tserkovnyak}}]{sato10}%
  \BibitemOpen
  \bibfield  {author} {\bibinfo {author} {\bibfnamefont {K.}~\bibnamefont
  {Sato}}, \bibinfo {author} {\bibfnamefont {D.}~\bibnamefont {Loss}}, \ and\
  \bibinfo {author} {\bibfnamefont {Y.}~\bibnamefont {Tserkovnyak}},\ }\href
  {\doibase 10.1103/PhysRevLett.105.226401} {\bibfield  {journal} {\bibinfo
  {journal} {Phys. Rev. Lett.}\ }\textbf {\bibinfo {volume} {105}},\ \bibinfo
  {pages} {226401} (\bibinfo {year} {2010})}\BibitemShut {NoStop}%
\bibitem [{\citenamefont {Hofstetter}\ \emph {et~al.}(2009)\citenamefont
  {Hofstetter}, \citenamefont {Csonka}, \citenamefont {Nyg\aa rd},\ and\
  \citenamefont {Sch\"onenberger}}]{hofstetter09}%
  \BibitemOpen
  \bibfield  {author} {\bibinfo {author} {\bibfnamefont {L.}~\bibnamefont
  {Hofstetter}}, \bibinfo {author} {\bibfnamefont {S.}~\bibnamefont {Csonka}},
  \bibinfo {author} {\bibfnamefont {J.}~\bibnamefont {Nyg\aa rd}}, \ and\
  \bibinfo {author} {\bibfnamefont {C.}~\bibnamefont {Sch\"onenberger}},\
  }\href {\doibase 10.1038/nature08432} {\bibfield  {journal} {\bibinfo
  {journal} {Nature (London)}\ }\textbf {\bibinfo {volume} {461}},\ \bibinfo
  {pages} {960} (\bibinfo {year} {2009})}\BibitemShut {NoStop}%
\bibitem [{\citenamefont {Herrmann}\ \emph {et~al.}(2010)\citenamefont
  {Herrmann}, \citenamefont {Portier}, \citenamefont {Roche}, \citenamefont
  {Levy~Yeyati}, \citenamefont {Kontos},\ and\ \citenamefont
  {Strunk}}]{herrmann10}%
  \BibitemOpen
  \bibfield  {author} {\bibinfo {author} {\bibfnamefont {L.~G.}\ \bibnamefont
  {Herrmann}}, \bibinfo {author} {\bibfnamefont {F.}~\bibnamefont {Portier}},
  \bibinfo {author} {\bibfnamefont {P.}~\bibnamefont {Roche}}, \bibinfo
  {author} {\bibfnamefont {A.}~\bibnamefont {Levy~Yeyati}}, \bibinfo {author}
  {\bibfnamefont {T.}~\bibnamefont {Kontos}}, \ and\ \bibinfo {author}
  {\bibfnamefont {C.}~\bibnamefont {Strunk}},\ }\href {\doibase
  10.1103/PhysRevLett.104.026801} {\bibfield  {journal} {\bibinfo  {journal}
  {Phys. Rev. Lett.}\ }\textbf {\bibinfo {volume} {104}},\ \bibinfo {pages}
  {026801} (\bibinfo {year} {2010})}\BibitemShut {NoStop}%
\bibitem [{\citenamefont {Das}\ \emph {et~al.}(2012)\citenamefont {Das},
  \citenamefont {Ronen}, \citenamefont {Heiblum}, \citenamefont {Mahalu},
  \citenamefont {Kretinin},\ and\ \citenamefont {Shtrikman}}]{das12}%
  \BibitemOpen
  \bibfield  {author} {\bibinfo {author} {\bibfnamefont {A.}~\bibnamefont
  {Das}}, \bibinfo {author} {\bibfnamefont {Y.}~\bibnamefont {Ronen}}, \bibinfo
  {author} {\bibfnamefont {M.}~\bibnamefont {Heiblum}}, \bibinfo {author}
  {\bibfnamefont {D.}~\bibnamefont {Mahalu}}, \bibinfo {author} {\bibfnamefont
  {A.~V.}\ \bibnamefont {Kretinin}}, \ and\ \bibinfo {author} {\bibfnamefont
  {H.}~\bibnamefont {Shtrikman}},\ }\href {\doibase 10.1038/ncomms2169}
  {\bibfield  {journal} {\bibinfo  {journal} {Nat. Commun.}\ }\textbf {\bibinfo
  {volume} {3}},\ \bibinfo {pages} {1165} (\bibinfo {year} {2012})}\BibitemShut
  {NoStop}%
\bibitem [{\citenamefont {Wei}\ and\ \citenamefont
  {Chandrasekhar}(2010)}]{wei10}%
  \BibitemOpen
  \bibfield  {author} {\bibinfo {author} {\bibfnamefont {J.}~\bibnamefont
  {Wei}}\ and\ \bibinfo {author} {\bibfnamefont {V.}~\bibnamefont
  {Chandrasekhar}},\ }\href {\doibase 10.1038/nphys1669} {\bibfield  {journal}
  {\bibinfo  {journal} {Nature Phys.}\ }\textbf {\bibinfo {volume} {6}},\
  \bibinfo {pages} {494} (\bibinfo {year} {2010})}\BibitemShut {NoStop}%
\bibitem [{\citenamefont {Kawabata}(2001)}]{kawabata01}%
  \BibitemOpen
  \bibfield  {author} {\bibinfo {author} {\bibfnamefont {S.}~\bibnamefont
  {Kawabata}},\ }\href {\doibase 10.1143/JPSJ.70.1210} {\bibfield  {journal}
  {\bibinfo  {journal} {J. Phys. Soc. Jpn.}\ }\textbf {\bibinfo {volume}
  {70}},\ \bibinfo {pages} {1210} (\bibinfo {year} {2001})}\BibitemShut
  {NoStop}%
\bibitem [{\citenamefont {Chtchelkatchev}\ \emph {et~al.}(2002)\citenamefont
  {Chtchelkatchev}, \citenamefont {Blatter}, \citenamefont {Lesovik},\ and\
  \citenamefont {Martin}}]{chtchelkatchev02}%
  \BibitemOpen
  \bibfield  {author} {\bibinfo {author} {\bibfnamefont {N.~M.}\ \bibnamefont
  {Chtchelkatchev}}, \bibinfo {author} {\bibfnamefont {G.}~\bibnamefont
  {Blatter}}, \bibinfo {author} {\bibfnamefont {G.~B.}\ \bibnamefont
  {Lesovik}}, \ and\ \bibinfo {author} {\bibfnamefont {T.}~\bibnamefont
  {Martin}},\ }\href {\doibase 10.1103/PhysRevB.66.161320} {\bibfield
  {journal} {\bibinfo  {journal} {Phys. Rev. B}\ }\textbf {\bibinfo {volume}
  {66}},\ \bibinfo {pages} {161320} (\bibinfo {year} {2002})}\BibitemShut
  {NoStop}%
\bibitem [{\citenamefont {Samuelsson}\ \emph {et~al.}(2003)\citenamefont
  {Samuelsson}, \citenamefont {Sukhorukov},\ and\ \citenamefont
  {B\"uttiker}}]{samuelsson03}%
  \BibitemOpen
  \bibfield  {author} {\bibinfo {author} {\bibfnamefont {P.}~\bibnamefont
  {Samuelsson}}, \bibinfo {author} {\bibfnamefont {E.~V.}\ \bibnamefont
  {Sukhorukov}}, \ and\ \bibinfo {author} {\bibfnamefont {M.}~\bibnamefont
  {B\"uttiker}},\ }\href {\doibase 10.1103/PhysRevLett.91.157002} {\bibfield
  {journal} {\bibinfo  {journal} {Phys. Rev. Lett.}\ }\textbf {\bibinfo
  {volume} {91}},\ \bibinfo {pages} {157002} (\bibinfo {year}
  {2003})}\BibitemShut {NoStop}%
\bibitem [{\citenamefont {Beenakker}\ \emph {et~al.}(2003)\citenamefont
  {Beenakker}, \citenamefont {Emary}, \citenamefont {Kindermann},\ and\
  \citenamefont {van Velsen}}]{beenakker03}%
  \BibitemOpen
  \bibfield  {author} {\bibinfo {author} {\bibfnamefont {C.~W.~J.}\
  \bibnamefont {Beenakker}}, \bibinfo {author} {\bibfnamefont {C.}~\bibnamefont
  {Emary}}, \bibinfo {author} {\bibfnamefont {M.}~\bibnamefont {Kindermann}}, \
  and\ \bibinfo {author} {\bibfnamefont {J.~L.}\ \bibnamefont {van Velsen}},\
  }\href {\doibase 10.1103/PhysRevLett.91.147901} {\bibfield  {journal}
  {\bibinfo  {journal} {Phys. Rev. Lett.}\ }\textbf {\bibinfo {volume} {91}},\
  \bibinfo {pages} {147901} (\bibinfo {year} {2003})}\BibitemShut {NoStop}%
\bibitem [{\citenamefont {Samuelsson}\ \emph
  {et~al.}(2004{\natexlab{a}})\citenamefont {Samuelsson}, \citenamefont
  {Sukhorukov},\ and\ \citenamefont {B\"uttiker}}]{samuelsson04-1}%
  \BibitemOpen
  \bibfield  {author} {\bibinfo {author} {\bibfnamefont {P.}~\bibnamefont
  {Samuelsson}}, \bibinfo {author} {\bibfnamefont {E.~V.}\ \bibnamefont
  {Sukhorukov}}, \ and\ \bibinfo {author} {\bibfnamefont {M.}~\bibnamefont
  {B\"uttiker}},\ }\href {\doibase 10.1103/PhysRevLett.92.026805} {\bibfield
  {journal} {\bibinfo  {journal} {Phys. Rev. Lett.}\ }\textbf {\bibinfo
  {volume} {92}},\ \bibinfo {pages} {026805} (\bibinfo {year}
  {2004}{\natexlab{a}})}\BibitemShut {NoStop}%
\bibitem [{\citenamefont {Sauret}\ \emph {et~al.}(2005)\citenamefont {Sauret},
  \citenamefont {Martin},\ and\ \citenamefont {Feinberg}}]{sauret05}%
  \BibitemOpen
  \bibfield  {author} {\bibinfo {author} {\bibfnamefont {O.}~\bibnamefont
  {Sauret}}, \bibinfo {author} {\bibfnamefont {T.}~\bibnamefont {Martin}}, \
  and\ \bibinfo {author} {\bibfnamefont {D.}~\bibnamefont {Feinberg}},\ }\href
  {\doibase 10.1103/PhysRevB.72.024544} {\bibfield  {journal} {\bibinfo
  {journal} {Phys. Rev. B}\ }\textbf {\bibinfo {volume} {72}},\ \bibinfo
  {pages} {024544} (\bibinfo {year} {2005})}\BibitemShut {NoStop}%
\bibitem [{\citenamefont {Chen}\ \emph {et~al.}(2012)\citenamefont {Chen},
  \citenamefont {Shen}, \citenamefont {Sheng}, \citenamefont {Wang},\ and\
  \citenamefont {Xing}}]{chen12}%
  \BibitemOpen
  \bibfield  {author} {\bibinfo {author} {\bibfnamefont {W.}~\bibnamefont
  {Chen}}, \bibinfo {author} {\bibfnamefont {R.}~\bibnamefont {Shen}}, \bibinfo
  {author} {\bibfnamefont {L.}~\bibnamefont {Sheng}}, \bibinfo {author}
  {\bibfnamefont {B.~G.}\ \bibnamefont {Wang}}, \ and\ \bibinfo {author}
  {\bibfnamefont {D.~Y.}\ \bibnamefont {Xing}},\ }\href {\doibase
  10.1103/PhysRevLett.109.036802} {\bibfield  {journal} {\bibinfo  {journal}
  {Phys. Rev. Lett.}\ }\textbf {\bibinfo {volume} {109}},\ \bibinfo {pages}
  {036802} (\bibinfo {year} {2012})}\BibitemShut {NoStop}%
\bibitem [{\citenamefont {Braunecker}\ \emph {et~al.}(2013)\citenamefont
  {Braunecker}, \citenamefont {Burset},\ and\ \citenamefont
  {Levy~Yeyati}}]{braunecker13}%
  \BibitemOpen
  \bibfield  {author} {\bibinfo {author} {\bibfnamefont {B.}~\bibnamefont
  {Braunecker}}, \bibinfo {author} {\bibfnamefont {P.}~\bibnamefont {Burset}},
  \ and\ \bibinfo {author} {\bibfnamefont {A.}~\bibnamefont {Levy~Yeyati}},\
  }\href {\doibase 10.1103/PhysRevLett.111.136806} {\bibfield  {journal}
  {\bibinfo  {journal} {Phys. Rev. Lett.}\ }\textbf {\bibinfo {volume} {111}},\
  \bibinfo {pages} {136806} (\bibinfo {year} {2013})}\BibitemShut {NoStop}%
\bibitem [{\citenamefont {Burkard}\ \emph {et~al.}(2000)\citenamefont
  {Burkard}, \citenamefont {Loss},\ and\ \citenamefont
  {Sukhorukov}}]{burkard00}%
  \BibitemOpen
  \bibfield  {author} {\bibinfo {author} {\bibfnamefont {G.}~\bibnamefont
  {Burkard}}, \bibinfo {author} {\bibfnamefont {D.}~\bibnamefont {Loss}}, \
  and\ \bibinfo {author} {\bibfnamefont {E.~V.}\ \bibnamefont {Sukhorukov}},\
  }\href {\doibase 10.1103/PhysRevB.61.R16303} {\bibfield  {journal} {\bibinfo
  {journal} {Phys. Rev. B}\ }\textbf {\bibinfo {volume} {61}},\ \bibinfo
  {pages} {R16303} (\bibinfo {year} {2000})}\BibitemShut {NoStop}%
\bibitem [{\citenamefont {Egues}\ \emph {et~al.}(2002)\citenamefont {Egues},
  \citenamefont {Burkard},\ and\ \citenamefont {Loss}}]{egues02}%
  \BibitemOpen
  \bibfield  {author} {\bibinfo {author} {\bibfnamefont {J.~C.}\ \bibnamefont
  {Egues}}, \bibinfo {author} {\bibfnamefont {G.}~\bibnamefont {Burkard}}, \
  and\ \bibinfo {author} {\bibfnamefont {D.}~\bibnamefont {Loss}},\ }\href
  {\doibase 10.1103/PhysRevLett.89.176401} {\bibfield  {journal} {\bibinfo
  {journal} {Phys. Rev. Lett.}\ }\textbf {\bibinfo {volume} {89}},\ \bibinfo
  {pages} {176401} (\bibinfo {year} {2002})}\BibitemShut {NoStop}%
\bibitem [{\citenamefont {Burkard}\ and\ \citenamefont
  {Loss}(2003)}]{burkard03}%
  \BibitemOpen
  \bibfield  {author} {\bibinfo {author} {\bibfnamefont {G.}~\bibnamefont
  {Burkard}}\ and\ \bibinfo {author} {\bibfnamefont {D.}~\bibnamefont {Loss}},\
  }\href {\doibase 10.1103/PhysRevLett.91.087903} {\bibfield  {journal}
  {\bibinfo  {journal} {Phys. Rev. Lett.}\ }\textbf {\bibinfo {volume} {91}},\
  \bibinfo {pages} {087903} (\bibinfo {year} {2003})}\BibitemShut {NoStop}%
\bibitem [{\citenamefont {Hu}\ and\ \citenamefont {Das~Sarma}(2004)}]{hu04}%
  \BibitemOpen
  \bibfield  {author} {\bibinfo {author} {\bibfnamefont {X.}~\bibnamefont
  {Hu}}\ and\ \bibinfo {author} {\bibfnamefont {S.}~\bibnamefont {Das~Sarma}},\
  }\href {\doibase 10.1103/PhysRevB.69.115312} {\bibfield  {journal} {\bibinfo
  {journal} {Phys. Rev. B}\ }\textbf {\bibinfo {volume} {69}},\ \bibinfo
  {pages} {115312} (\bibinfo {year} {2004})}\BibitemShut {NoStop}%
\bibitem [{\citenamefont {Samuelsson}\ \emph
  {et~al.}(2004{\natexlab{b}})\citenamefont {Samuelsson}, \citenamefont
  {Sukhorukov},\ and\ \citenamefont {B\"uttiker}}]{samuelsson04-2}%
  \BibitemOpen
  \bibfield  {author} {\bibinfo {author} {\bibfnamefont {P.}~\bibnamefont
  {Samuelsson}}, \bibinfo {author} {\bibfnamefont {E.~V.}\ \bibnamefont
  {Sukhorukov}}, \ and\ \bibinfo {author} {\bibfnamefont {M.}~\bibnamefont
  {B\"uttiker}},\ }\href {\doibase 10.1103/PhysRevB.70.115330} {\bibfield
  {journal} {\bibinfo  {journal} {Phys. Rev. B}\ }\textbf {\bibinfo {volume}
  {70}},\ \bibinfo {pages} {115330} (\bibinfo {year}
  {2004}{\natexlab{b}})}\BibitemShut {NoStop}%
\bibitem [{\citenamefont {Egues}\ \emph {et~al.}(2005)\citenamefont {Egues},
  \citenamefont {Burkard}, \citenamefont {Saraga}, \citenamefont {Schliemann},\
  and\ \citenamefont {Loss}}]{egues05}%
  \BibitemOpen
  \bibfield  {author} {\bibinfo {author} {\bibfnamefont {J.~C.}\ \bibnamefont
  {Egues}}, \bibinfo {author} {\bibfnamefont {G.}~\bibnamefont {Burkard}},
  \bibinfo {author} {\bibfnamefont {D.~S.}\ \bibnamefont {Saraga}}, \bibinfo
  {author} {\bibfnamefont {J.}~\bibnamefont {Schliemann}}, \ and\ \bibinfo
  {author} {\bibfnamefont {D.}~\bibnamefont {Loss}},\ }\href {\doibase
  10.1103/PhysRevB.72.235326} {\bibfield  {journal} {\bibinfo  {journal} {Phys.
  Rev. B}\ }\textbf {\bibinfo {volume} {72}},\ \bibinfo {pages} {235326}
  (\bibinfo {year} {2005})}\BibitemShut {NoStop}%
\bibitem [{\citenamefont {Giovannetti}\ \emph {et~al.}(2006)\citenamefont
  {Giovannetti}, \citenamefont {Frustaglia}, \citenamefont {Taddei},\ and\
  \citenamefont {Fazio}}]{giovannetti06}%
  \BibitemOpen
  \bibfield  {author} {\bibinfo {author} {\bibfnamefont {V.}~\bibnamefont
  {Giovannetti}}, \bibinfo {author} {\bibfnamefont {D.}~\bibnamefont
  {Frustaglia}}, \bibinfo {author} {\bibfnamefont {F.}~\bibnamefont {Taddei}},
  \ and\ \bibinfo {author} {\bibfnamefont {R.}~\bibnamefont {Fazio}},\ }\href
  {\doibase 10.1103/PhysRevB.74.115315} {\bibfield  {journal} {\bibinfo
  {journal} {Phys. Rev. B}\ }\textbf {\bibinfo {volume} {74}},\ \bibinfo
  {pages} {115315} (\bibinfo {year} {2006})}\BibitemShut {NoStop}%
\bibitem [{\citenamefont {San-Jose}\ and\ \citenamefont
  {Prada}(2006)}]{prada06}%
  \BibitemOpen
  \bibfield  {author} {\bibinfo {author} {\bibfnamefont {P.}~\bibnamefont
  {San-Jose}}\ and\ \bibinfo {author} {\bibfnamefont {E.}~\bibnamefont
  {Prada}},\ }\href {\doibase 10.1103/PhysRevB.74.045305} {\bibfield  {journal}
  {\bibinfo  {journal} {Phys. Rev. B}\ }\textbf {\bibinfo {volume} {74}},\
  \bibinfo {pages} {045305} (\bibinfo {year} {2006})}\BibitemShut {NoStop}%
\bibitem [{\citenamefont {Mazza}\ \emph {et~al.}(2013)\citenamefont {Mazza},
  \citenamefont {Braunecker}, \citenamefont {Recher},\ and\ \citenamefont
  {Levy~Yeyati}}]{mazza13}%
  \BibitemOpen
  \bibfield  {author} {\bibinfo {author} {\bibfnamefont {F.}~\bibnamefont
  {Mazza}}, \bibinfo {author} {\bibfnamefont {B.}~\bibnamefont {Braunecker}},
  \bibinfo {author} {\bibfnamefont {P.}~\bibnamefont {Recher}}, \ and\ \bibinfo
  {author} {\bibfnamefont {A.}~\bibnamefont {Levy~Yeyati}},\ }\href {\doibase
  10.1103/PhysRevB.88.195403} {\bibfield  {journal} {\bibinfo  {journal} {Phys.
  Rev. B}\ }\textbf {\bibinfo {volume} {88}},\ \bibinfo {pages} {195403}
  (\bibinfo {year} {2013})}\BibitemShut {NoStop}%
\bibitem [{\citenamefont {Schroer}\ \emph {et~al.}(2014)\citenamefont
  {Schroer}, \citenamefont {Braunecker}, \citenamefont {Levy~Yeyati},\ and\
  \citenamefont {Recher}}]{schroer14}%
  \BibitemOpen
  \bibfield  {author} {\bibinfo {author} {\bibfnamefont {A.}~\bibnamefont
  {Schroer}}, \bibinfo {author} {\bibfnamefont {B.}~\bibnamefont {Braunecker}},
  \bibinfo {author} {\bibfnamefont {A.}~\bibnamefont {Levy~Yeyati}}, \ and\
  \bibinfo {author} {\bibfnamefont {P.}~\bibnamefont {Recher}},\ }\href 
  {\doibase 10.1103/PhysRevLett.113.266401} {\bibfield  {journal}
  {\bibinfo  {journal} {Phys. Rev. Lett.}\ }\textbf {\bibinfo {volume} {113}},\
  \bibinfo {pages} {266401} (\bibinfo {year} {2014})}\BibitemShut {NoStop}%
\bibitem [{\citenamefont {Freedman}\ and\ \citenamefont
  {Clauser}(1972)}]{freedman72}%
  \BibitemOpen
  \bibfield  {author} {\bibinfo {author} {\bibfnamefont {S.~J.}\ \bibnamefont
  {Freedman}}\ and\ \bibinfo {author} {\bibfnamefont {J.~F.}\ \bibnamefont
  {Clauser}},\ }\href {\doibase 10.1103/PhysRevLett.28.938} {\bibfield
  {journal} {\bibinfo  {journal} {Phys. Rev. Lett.}\ }\textbf {\bibinfo
  {volume} {28}},\ \bibinfo {pages} {938} (\bibinfo {year} {1972})}\BibitemShut
  {NoStop}%
\bibitem [{\citenamefont {Aspect}\ \emph {et~al.}(1982)\citenamefont {Aspect},
  \citenamefont {Dalibard},\ and\ \citenamefont {Roger}}]{aspect82}%
  \BibitemOpen
  \bibfield  {author} {\bibinfo {author} {\bibfnamefont {A.}~\bibnamefont
  {Aspect}}, \bibinfo {author} {\bibfnamefont {J.}~\bibnamefont {Dalibard}}, \
  and\ \bibinfo {author} {\bibfnamefont {G.}~\bibnamefont {Roger}},\ }\href
  {\doibase 10.1103/PhysRevLett.49.1804} {\bibfield  {journal} {\bibinfo
  {journal} {Phys. Rev. Lett.}\ }\textbf {\bibinfo {volume} {49}},\ \bibinfo
  {pages} {1804} (\bibinfo {year} {1982})}\BibitemShut {NoStop}%
\bibitem [{\citenamefont {Weihs}\ \emph {et~al.}(1998)\citenamefont {Weihs},
  \citenamefont {Jennewein}, \citenamefont {Simon}, \citenamefont
  {Weinfurter},\ and\ \citenamefont {Zeilinger}}]{weihs98}%
  \BibitemOpen
  \bibfield  {author} {\bibinfo {author} {\bibfnamefont {G.}~\bibnamefont
  {Weihs}}, \bibinfo {author} {\bibfnamefont {T.}~\bibnamefont {Jennewein}},
  \bibinfo {author} {\bibfnamefont {C.}~\bibnamefont {Simon}}, \bibinfo
  {author} {\bibfnamefont {H.}~\bibnamefont {Weinfurter}}, \ and\ \bibinfo
  {author} {\bibfnamefont {A.}~\bibnamefont {Zeilinger}},\ }\href {\doibase
  10.1103/PhysRevLett.81.5039} {\bibfield  {journal} {\bibinfo  {journal}
  {Phys. Rev. Lett.}\ }\textbf {\bibinfo {volume} {81}},\ \bibinfo {pages}
  {5039} (\bibinfo {year} {1998})}\BibitemShut {NoStop}%
\bibitem [{\citenamefont {Bell}(1964)}]{bell64}%
  \BibitemOpen
  \bibfield  {author} {\bibinfo {author} {\bibfnamefont {J.~S.}\ \bibnamefont
  {Bell}},\ }\href@noop {} {\bibfield  {journal} {\bibinfo  {journal}
  {Physics}\ }\textbf {\bibinfo {volume} {1}} (\bibinfo {year}
  {1964})}\BibitemShut {NoStop}%
\bibitem [{\citenamefont {Clauser}\ \emph {et~al.}(1969)\citenamefont
  {Clauser}, \citenamefont {Horne}, \citenamefont {Shimony},\ and\
  \citenamefont {Holt}}]{clauser69}%
  \BibitemOpen
  \bibfield  {author} {\bibinfo {author} {\bibfnamefont {J.~F.}\ \bibnamefont
  {Clauser}}, \bibinfo {author} {\bibfnamefont {M.~A.}\ \bibnamefont {Horne}},
  \bibinfo {author} {\bibfnamefont {A.}~\bibnamefont {Shimony}}, \ and\
  \bibinfo {author} {\bibfnamefont {R.~A.}\ \bibnamefont {Holt}},\ }\href
  {\doibase 10.1103/PhysRevLett.23.880} {\bibfield  {journal} {\bibinfo
  {journal} {Phys. Rev. Lett.}\ }\textbf {\bibinfo {volume} {23}},\ \bibinfo
  {pages} {880} (\bibinfo {year} {1969})}\BibitemShut {NoStop}%
\bibitem [{\citenamefont {Hanamura}(2002)}]{hanamura02}%
  \BibitemOpen
  \bibfield  {author} {\bibinfo {author} {\bibfnamefont {E.}~\bibnamefont
  {Hanamura}},\ }\href {\doibase
  10.1002/1521-3951(200211)234:1<166::AID-PSSB166>3.0.CO;2-J} {\bibfield
  {journal} {\bibinfo  {journal} {Phys. Status Solidi B}\ }\textbf {\bibinfo
  {volume} {234}},\ \bibinfo {pages} {166} (\bibinfo {year}
  {2002})}\BibitemShut {NoStop}%
\bibitem [{\citenamefont {Asano}\ \emph {et~al.}(2009)\citenamefont {Asano},
  \citenamefont {Suemune}, \citenamefont {Takayanagi},\ and\ \citenamefont
  {Hanamura}}]{asano09}%
  \BibitemOpen
  \bibfield  {author} {\bibinfo {author} {\bibfnamefont {Y.}~\bibnamefont
  {Asano}}, \bibinfo {author} {\bibfnamefont {I.}~\bibnamefont {Suemune}},
  \bibinfo {author} {\bibfnamefont {H.}~\bibnamefont {Takayanagi}}, \ and\
  \bibinfo {author} {\bibfnamefont {E.}~\bibnamefont {Hanamura}},\ }\href
  {\doibase 10.1103/PhysRevLett.103.187001} {\bibfield  {journal} {\bibinfo
  {journal} {Phys. Rev. Lett.}\ }\textbf {\bibinfo {volume} {103}},\ \bibinfo
  {pages} {187001} (\bibinfo {year} {2009})}\BibitemShut {NoStop}%
\bibitem [{\citenamefont {Sasakura}\ \emph {et~al.}(2011)\citenamefont
  {Sasakura}, \citenamefont {Kuramitsu}, \citenamefont {Hayashi}, \citenamefont
  {Tanaka}, \citenamefont {Akazaki}, \citenamefont {Hanamura}, \citenamefont
  {Inoue}, \citenamefont {Takayanagi}, \citenamefont {Asano}, \citenamefont
  {Hermannst\"adter}, \citenamefont {Kumano},\ and\ \citenamefont
  {Suemune}}]{sasakura11}%
  \BibitemOpen
  \bibfield  {author} {\bibinfo {author} {\bibfnamefont {H.}~\bibnamefont
  {Sasakura}}, \bibinfo {author} {\bibfnamefont {S.}~\bibnamefont {Kuramitsu}},
  \bibinfo {author} {\bibfnamefont {Y.}~\bibnamefont {Hayashi}}, \bibinfo
  {author} {\bibfnamefont {K.}~\bibnamefont {Tanaka}}, \bibinfo {author}
  {\bibfnamefont {T.}~\bibnamefont {Akazaki}}, \bibinfo {author} {\bibfnamefont
  {E.}~\bibnamefont {Hanamura}}, \bibinfo {author} {\bibfnamefont
  {R.}~\bibnamefont {Inoue}}, \bibinfo {author} {\bibfnamefont
  {H.}~\bibnamefont {Takayanagi}}, \bibinfo {author} {\bibfnamefont
  {Y.}~\bibnamefont {Asano}}, \bibinfo {author} {\bibfnamefont
  {C.}~\bibnamefont {Hermannst\"adter}}, \bibinfo {author} {\bibfnamefont
  {H.}~\bibnamefont {Kumano}}, \ and\ \bibinfo {author} {\bibfnamefont
  {I.}~\bibnamefont {Suemune}},\ }\href {\doibase
  10.1103/PhysRevLett.107.157403} {\bibfield  {journal} {\bibinfo  {journal}
  {Phys. Rev. Lett.}\ }\textbf {\bibinfo {volume} {107}},\ \bibinfo {pages}
  {157403} (\bibinfo {year} {2011})}\BibitemShut {NoStop}%
\bibitem [{\citenamefont {Suemune}\ \emph {et~al.}(2006)\citenamefont
  {Suemune}, \citenamefont {Akazaki}, \citenamefont {Tanaka}, \citenamefont
  {Jo}, \citenamefont {Uesugi}, \citenamefont {Endo}, \citenamefont {Kumano},
  \citenamefont {Hanamura}, \citenamefont {Takayanagi}, \citenamefont
  {Yamanishi},\ and\ \citenamefont {Kan}}]{suemune06}%
  \BibitemOpen
  \bibfield  {author} {\bibinfo {author} {\bibfnamefont {I.}~\bibnamefont
  {Suemune}}, \bibinfo {author} {\bibfnamefont {T.}~\bibnamefont {Akazaki}},
  \bibinfo {author} {\bibfnamefont {K.}~\bibnamefont {Tanaka}}, \bibinfo
  {author} {\bibfnamefont {M.}~\bibnamefont {Jo}}, \bibinfo {author}
  {\bibfnamefont {K.}~\bibnamefont {Uesugi}}, \bibinfo {author} {\bibfnamefont
  {M.}~\bibnamefont {Endo}}, \bibinfo {author} {\bibfnamefont {H.}~\bibnamefont
  {Kumano}}, \bibinfo {author} {\bibfnamefont {E.}~\bibnamefont {Hanamura}},
  \bibinfo {author} {\bibfnamefont {H.}~\bibnamefont {Takayanagi}}, \bibinfo
  {author} {\bibfnamefont {M.}~\bibnamefont {Yamanishi}}, \ and\ \bibinfo
  {author} {\bibfnamefont {H.}~\bibnamefont {Kan}},\ }\href {\doibase
  10.1143/JJAP.45.9264} {\bibfield  {journal} {\bibinfo  {journal} {Jpn. J.
  Appl. Phys.}\ }\textbf {\bibinfo {volume} {45}},\ \bibinfo {pages} {9264}
  (\bibinfo {year} {2006})}\BibitemShut {NoStop}%
\bibitem [{\citenamefont {Recher}\ \emph {et~al.}(2010)\citenamefont {Recher},
  \citenamefont {Nazarov},\ and\ \citenamefont {Kouwenhoven}}]{recher10}%
  \BibitemOpen
  \bibfield  {author} {\bibinfo {author} {\bibfnamefont {P.}~\bibnamefont
  {Recher}}, \bibinfo {author} {\bibfnamefont {Y.~V.}\ \bibnamefont {Nazarov}},
  \ and\ \bibinfo {author} {\bibfnamefont {L.~P.}\ \bibnamefont
  {Kouwenhoven}},\ }\href {\doibase 10.1103/PhysRevLett.104.156802} {\bibfield
  {journal} {\bibinfo  {journal} {Phys. Rev. Lett.}\ }\textbf {\bibinfo
  {volume} {104}},\ \bibinfo {pages} {156802} (\bibinfo {year}
  {2010})}\BibitemShut {NoStop}%
\bibitem [{\citenamefont {Hassler}\ \emph {et~al.}(2010)\citenamefont
  {Hassler}, \citenamefont {Nazarov},\ and\ \citenamefont
  {Kouwenhoven}}]{hassler10}%
  \BibitemOpen
  \bibfield  {author} {\bibinfo {author} {\bibfnamefont {F.}~\bibnamefont
  {Hassler}}, \bibinfo {author} {\bibfnamefont {Y.~V.}\ \bibnamefont
  {Nazarov}}, \ and\ \bibinfo {author} {\bibfnamefont {L.~P.}\ \bibnamefont
  {Kouwenhoven}},\ }\href {\doibase 10.1088/0957-4484/21/27/274004} {\bibfield
  {journal} {\bibinfo  {journal} {Nanotechnology}\ }\textbf {\bibinfo {volume}
  {21}},\ \bibinfo {pages} {274004} (\bibinfo {year} {2010})}\BibitemShut
  {NoStop}%
\bibitem [{\citenamefont {Baireuther}\ \emph {et~al.}(2014)\citenamefont
  {Baireuther}, \citenamefont {Orth}, \citenamefont {Vekhter},\ and\
  \citenamefont {Schmalian}}]{baireuther14}%
  \BibitemOpen
  \bibfield  {author} {\bibinfo {author} {\bibfnamefont {P.}~\bibnamefont
  {Baireuther}}, \bibinfo {author} {\bibfnamefont {P.~P.}\ \bibnamefont
  {Orth}}, \bibinfo {author} {\bibfnamefont {I.}~\bibnamefont {Vekhter}}, \
  and\ \bibinfo {author} {\bibfnamefont {J.}~\bibnamefont {Schmalian}},\ }\href
  {\doibase 10.1103/PhysRevLett.112.077003} {\bibfield  {journal} {\bibinfo
  {journal} {Phys. Rev. Lett.}\ }\textbf {\bibinfo {volume} {112}},\ \bibinfo
  {pages} {077003} (\bibinfo {year} {2014})}\BibitemShut {NoStop}%
\bibitem [{\citenamefont {Hayat}\ \emph {et~al.}(2014)\citenamefont {Hayat},
  \citenamefont {Kee}, \citenamefont {Burch},\ and\ \citenamefont
  {Steinberg}}]{hayat14}%
  \BibitemOpen
  \bibfield  {author} {\bibinfo {author} {\bibfnamefont {A.}~\bibnamefont
  {Hayat}}, \bibinfo {author} {\bibfnamefont {H.-Y.}\ \bibnamefont {Kee}},
  \bibinfo {author} {\bibfnamefont {K.~S.}\ \bibnamefont {Burch}}, \ and\
  \bibinfo {author} {\bibfnamefont {A.~M.}\ \bibnamefont {Steinberg}},\ }\href
  {\doibase 10.1103/PhysRevB.89.094508} {\bibfield  {journal} {\bibinfo
  {journal} {Phys. Rev. B}\ }\textbf {\bibinfo {volume} {89}},\ \bibinfo
  {pages} {094508} (\bibinfo {year} {2014})}\BibitemShut {NoStop}%
\bibitem [{\citenamefont {Godschalk}\ \emph {et~al.}(2011)\citenamefont
  {Godschalk}, \citenamefont {Hassler},\ and\ \citenamefont
  {Nazarov}}]{godschalk11}%
  \BibitemOpen
  \bibfield  {author} {\bibinfo {author} {\bibfnamefont {F.}~\bibnamefont
  {Godschalk}}, \bibinfo {author} {\bibfnamefont {F.}~\bibnamefont {Hassler}},
  \ and\ \bibinfo {author} {\bibfnamefont {Y.~V.}\ \bibnamefont {Nazarov}},\
  }\href {\doibase 10.1103/PhysRevLett.107.073901} {\bibfield  {journal}
  {\bibinfo  {journal} {Phys. Rev. Lett.}\ }\textbf {\bibinfo {volume} {107}},\
  \bibinfo {pages} {073901} (\bibinfo {year} {2011})}\BibitemShut {NoStop}%
\bibitem [{\citenamefont {Godschalk}\ and\ \citenamefont
  {Nazarov}(2013)}]{godschalk13}%
  \BibitemOpen
  \bibfield  {author} {\bibinfo {author} {\bibfnamefont {F.}~\bibnamefont
  {Godschalk}}\ and\ \bibinfo {author} {\bibfnamefont {Y.~V.}\ \bibnamefont
  {Nazarov}},\ }\href {\doibase 10.1103/PhysRevB.87.094511} {\bibfield
  {journal} {\bibinfo  {journal} {Phys. Rev. B}\ }\textbf {\bibinfo {volume}
  {87}},\ \bibinfo {pages} {094511} (\bibinfo {year} {2013})}\BibitemShut
  {NoStop}%
\bibitem [{\citenamefont {Godschalk}\ and\ \citenamefont
  {Nazarov}(2014)}]{godschalk14}%
  \BibitemOpen
  \bibfield  {author} {\bibinfo {author} {\bibfnamefont {F.}~\bibnamefont
  {Godschalk}}\ and\ \bibinfo {author} {\bibfnamefont {Y.~V.}\ \bibnamefont
  {Nazarov}},\ }\href {\doibase 10.1103/PhysRevB.89.104502} {\bibfield
  {journal} {\bibinfo  {journal} {Phys. Rev. B}\ }\textbf {\bibinfo {volume}
  {89}},\ \bibinfo {pages} {104502} (\bibinfo {year} {2014})}\BibitemShut
  {NoStop}%
\bibitem [{\citenamefont {Gywat}\ \emph {et~al.}(2002)\citenamefont {Gywat},
  \citenamefont {Burkard},\ and\ \citenamefont {Loss}}]{gywat02}%
  \BibitemOpen
  \bibfield  {author} {\bibinfo {author} {\bibfnamefont {O.}~\bibnamefont
  {Gywat}}, \bibinfo {author} {\bibfnamefont {G.}~\bibnamefont {Burkard}}, \
  and\ \bibinfo {author} {\bibfnamefont {D.}~\bibnamefont {Loss}},\ }\href
  {\doibase 10.1103/PhysRevB.65.205329} {\bibfield  {journal} {\bibinfo
  {journal} {Phys. Rev. B}\ }\textbf {\bibinfo {volume} {65}},\ \bibinfo
  {pages} {205329} (\bibinfo {year} {2002})}\BibitemShut {NoStop}%
\bibitem [{\citenamefont {Cerletti}\ \emph {et~al.}(2005)\citenamefont
  {Cerletti}, \citenamefont {Gywat},\ and\ \citenamefont {Loss}}]{cerletti05}%
  \BibitemOpen
  \bibfield  {author} {\bibinfo {author} {\bibfnamefont {V.}~\bibnamefont
  {Cerletti}}, \bibinfo {author} {\bibfnamefont {O.}~\bibnamefont {Gywat}}, \
  and\ \bibinfo {author} {\bibfnamefont {D.}~\bibnamefont {Loss}},\ }\href
  {\doibase 10.1103/PhysRevB.72.115316} {\bibfield  {journal} {\bibinfo
  {journal} {Phys. Rev. B}\ }\textbf {\bibinfo {volume} {72}},\ \bibinfo
  {pages} {115316} (\bibinfo {year} {2005})}\BibitemShut {NoStop}%
\bibitem [{\citenamefont {Budich}\ and\ \citenamefont
  {Trauzettel}(2010)}]{budich09}%
  \BibitemOpen
  \bibfield  {author} {\bibinfo {author} {\bibfnamefont {J.~C.}\ \bibnamefont
  {Budich}}\ and\ \bibinfo {author} {\bibfnamefont {B.}~\bibnamefont
  {Trauzettel}},\ }\href {\doibase 10.1088/0957-4484/21/27/274001} {\bibfield
  {journal} {\bibinfo  {journal} {Nanotechnology}\ }\textbf {\bibinfo {volume}
  {21}},\ \bibinfo {pages} {274001} (\bibinfo {year} {2010})}\BibitemShut
  {NoStop}%
\bibitem [{\citenamefont {Nigg}\ \emph {et~al.}(2014)\citenamefont {Nigg},
  \citenamefont {Tiwari}, \citenamefont {Walter},\ and\ \citenamefont
  {Schmidt}}]{nigg14}%
  \BibitemOpen
  \bibfield  {author} {\bibinfo {author} {\bibfnamefont {S.~E.}\ \bibnamefont
  {Nigg}}, \bibinfo {author} {\bibfnamefont {R.~P.}\ \bibnamefont {Tiwari}},
  \bibinfo {author} {\bibfnamefont {S.}~\bibnamefont {Walter}}, \ and\ \bibinfo
  {author} {\bibfnamefont {T.~L.}\ \bibnamefont {Schmidt}},\ }\href@noop {} {
  \Eprint {http://arxiv.org/abs/arxiv:1411.3945}}
  {arxiv:1411.3945} \BibitemShut {NoStop}%
\bibitem [{\citenamefont {Vogel}\ \emph {et~al.}(2001)\citenamefont {Vogel},
  \citenamefont {Welsch},\ and\ \citenamefont {Wallentowitz}}]{vogel01}%
  \BibitemOpen
  \bibfield  {author} {\bibinfo {author} {\bibfnamefont {W.}~\bibnamefont
  {Vogel}}, \bibinfo {author} {\bibfnamefont {D.-G.}\ \bibnamefont {Welsch}}, \
  and\ \bibinfo {author} {\bibfnamefont {S.}~\bibnamefont {Wallentowitz}},\
  }\href@noop {} {\emph {\bibinfo {title} {Quantum Optics An Introduction}}}\
  (\bibinfo  {publisher} {WILEY-VCH Verlag, Berlin},\ \bibinfo {year}
  {2001})\BibitemShut {NoStop}%
\bibitem [{\citenamefont {Vivoli}\ \emph {et~al.}(2014)\citenamefont {Vivoli},
  \citenamefont {Sekatski}, \citenamefont {Bancal}, \citenamefont {Lim},
  \citenamefont {Christensen}, \citenamefont {Martin}, \citenamefont {Thew},
  \citenamefont {Zbinden}, \citenamefont {Gisin},\ and\ \citenamefont
  {Sangouard}}]{vivoli14}%
  \BibitemOpen
  \bibfield  {author} {\bibinfo {author} {\bibfnamefont {V.~C.}\ \bibnamefont
  {Vivoli}}, \bibinfo {author} {\bibfnamefont {P.}~\bibnamefont {Sekatski}},
  \bibinfo {author} {\bibfnamefont {J.~D.}\ \bibnamefont {Bancal}}, \bibinfo
  {author} {\bibfnamefont {C.~C.~W.}\ \bibnamefont {Lim}}, \bibinfo {author}
  {\bibfnamefont {B.~G.}\ \bibnamefont {Christensen}}, \bibinfo {author}
  {\bibfnamefont {A.}~\bibnamefont {Martin}}, \bibinfo {author} {\bibfnamefont
  {R.~T.}\ \bibnamefont {Thew}}, \bibinfo {author} {\bibfnamefont
  {H.}~\bibnamefont {Zbinden}}, \bibinfo {author} {\bibfnamefont
  {N.}~\bibnamefont {Gisin}}, \ and\ \bibinfo {author} {\bibfnamefont
  {N.}~\bibnamefont {Sangouard}},\ }\href {http://www.arxiv.org/abs/1405.1939/}
  {\bibinfo {note} {arxiv:1405.1939}}\BibitemShut
  {NoStop}%
\bibitem [{sup()}]{suppl}%
  \BibitemOpen
  \href@noop {} {\bibinfo  {journal} {See Supplemental Material for
  further details}\ }\BibitemShut {NoStop}%
\bibitem [{Note1()}]{Note1}%
  \BibitemOpen
\bibfield  {journal} {  }\bibinfo {note} {For this estimate, we assume a
  worst-case set of parameters: a high quality factor $Q=10^6$, infrared light
  $f=10$~THz and a short coincidence interval $\Delta t=1$~ps.}
  \BibitemShut {Stop}%
\bibitem [{\citenamefont {Gardiner}\ and\ \citenamefont
  {Zoller}(2000)}]{gardiner00}%
  \BibitemOpen
  \bibfield  {author} {\bibinfo {author} {\bibfnamefont {C.~W.}\ \bibnamefont
  {Gardiner}}\ and\ \bibinfo {author} {\bibfnamefont {P.}~\bibnamefont
  {Zoller}},\ }\href@noop {} {\emph {\bibinfo {title} {Quantum Noise}}}\
  (\bibinfo  {publisher} {Springer, Berlin},\ \bibinfo {year}
  {2000})\BibitemShut {NoStop}%
\bibitem [{Note2()}]{Note2}%
  \BibitemOpen
  \bibinfo {note} {At finite Coulomb repulsion $U<|\Delta |$, the electron-hole
  system can be diagonalized perturbatively in $1/U$ or numerically. This
  substantially complicates the notation while not producing any new effects,
  so we focus on the (experimentally more relevant) case $U\gg |\Delta |$, in
  which ECT between the QDs via the SC lead is the dominant source of local
  photon pairs.}\BibitemShut {Stop}%
\bibitem [{\citenamefont {Titov}\ \emph {et~al.}(2005)\citenamefont {Titov},
  \citenamefont {Trauzettel}, \citenamefont {Michaelis},\ and\ \citenamefont
  {Beenakker}}]{titov05}%
  \BibitemOpen
  \bibfield  {author} {\bibinfo {author} {\bibfnamefont {M.}~\bibnamefont
  {Titov}}, \bibinfo {author} {\bibfnamefont {B.}~\bibnamefont {Trauzettel}},
  \bibinfo {author} {\bibfnamefont {B.}~\bibnamefont {Michaelis}}, \ and\
  \bibinfo {author} {\bibfnamefont {C.~W.~J.}\ \bibnamefont {Beenakker}},\
  }\href {http://stacks.iop.org/1367-2630/7/i=1/a=186} {\bibfield  {journal}
  {\bibinfo  {journal} {New Journal of Physics}\ }\textbf {\bibinfo {volume}
  {7}},\ \bibinfo {pages} {186} (\bibinfo {year} {2005})}\BibitemShut {NoStop}%
\bibitem [{Note3()}]{Note3}%
  \BibitemOpen
  \bibinfo {note} {Because of parity conservation we expect the photons to be
  entangled even if the cavity linewidth $\kappa $ is larger than $\Delta
  E^{\mu \nu }$ (but smaller than $\Delta $ in the SC leads and $U$). This,
  however, leads to a different regime in which single photons can be emitted
  sequentially, and requires a separate calculation.}\BibitemShut {Stop}%
\bibitem [{\citenamefont {Hennessy}\ \emph {et~al.}(2007)\citenamefont
  {Hennessy}, \citenamefont {Badolato}, \citenamefont {Winger}, \citenamefont
  {Gerace}, \citenamefont {Atature}, \citenamefont {Gulde}, \citenamefont
  {Falt}, \citenamefont {Hu},\ and\ \citenamefont {Imamoglu}}]{hennessy07}%
  \BibitemOpen
  \bibfield  {author} {\bibinfo {author} {\bibfnamefont {K.}~\bibnamefont
  {Hennessy}}, \bibinfo {author} {\bibfnamefont {A.}~\bibnamefont {Badolato}},
  \bibinfo {author} {\bibfnamefont {M.}~\bibnamefont {Winger}}, \bibinfo
  {author} {\bibfnamefont {D.}~\bibnamefont {Gerace}}, \bibinfo {author}
  {\bibfnamefont {M.}~\bibnamefont {Atature}}, \bibinfo {author} {\bibfnamefont
  {S.}~\bibnamefont {Gulde}}, \bibinfo {author} {\bibfnamefont
  {S.}~\bibnamefont {Falt}}, \bibinfo {author} {\bibfnamefont {E.~L.}\
  \bibnamefont {Hu}}, \ and\ \bibinfo {author} {\bibfnamefont {A.}~\bibnamefont
  {Imamoglu}},\ }\href {http://dx.doi.org/10.1038/nature05586} {\bibfield
  {journal} {\bibinfo  {journal} {Nature (London)}\ }\textbf {\bibinfo {volume}
  {445}},\ \bibinfo {pages} {896} (\bibinfo {year} {2007})}\BibitemShut
  {NoStop}%
\bibitem [{\citenamefont {Srinivasan}\ and\ \citenamefont
  {Painter}(2007)}]{srinivasan07}%
  \BibitemOpen
  \bibfield  {author} {\bibinfo {author} {\bibfnamefont {K.}~\bibnamefont
  {Srinivasan}}\ and\ \bibinfo {author} {\bibfnamefont {O.}~\bibnamefont
  {Painter}},\ }\href {http://dx.doi.org/10.1038/nature06274} {\bibfield
  {journal} {\bibinfo  {journal} {Nature}\ }\textbf {\bibinfo {volume} {450}},\
  \bibinfo {pages} {862} (\bibinfo {year} {2007})}\BibitemShut {NoStop}%
\bibitem [{\citenamefont {Loo}\ \emph {et~al.}(2010)\citenamefont {Loo},
  \citenamefont {Lanco}, \citenamefont {Lemaître}, \citenamefont {Sagnes},
  \citenamefont {Krebs}, \citenamefont {Voisin},\ and\ \citenamefont
  {Senellart}}]{loo10}%
  \BibitemOpen
  \bibfield  {author} {\bibinfo {author} {\bibfnamefont {V.}~\bibnamefont
  {Loo}}, \bibinfo {author} {\bibfnamefont {L.}~\bibnamefont {Lanco}}, \bibinfo
  {author} {\bibfnamefont {A.}~\bibnamefont {Lemaître}}, \bibinfo {author}
  {\bibfnamefont {I.}~\bibnamefont {Sagnes}}, \bibinfo {author} {\bibfnamefont
  {O.}~\bibnamefont {Krebs}}, \bibinfo {author} {\bibfnamefont
  {P.}~\bibnamefont {Voisin}}, \ and\ \bibinfo {author} {\bibfnamefont
  {P.}~\bibnamefont {Senellart}},\ }\href {\doibase
  http://dx.doi.org/10.1063/1.3527930} {\bibfield  {journal} {\bibinfo
  {journal} {Appl. Phys. Lett.}\ }\textbf {\bibinfo {volume} {97}},\ \bibinfo
  {eid} {241110} (\bibinfo {year} {2010})}\BibitemShut {NoStop}%
\bibitem [{\citenamefont {Ohta}\ \emph {et~al.}(2011)\citenamefont {Ohta},
  \citenamefont {Ota}, \citenamefont {Nomura}, \citenamefont {Kumagai},
  \citenamefont {Ishida}, \citenamefont {Iwamoto},\ and\ \citenamefont
  {Arakawa}}]{ohta11}%
  \BibitemOpen
  \bibfield  {author} {\bibinfo {author} {\bibfnamefont {R.}~\bibnamefont
  {Ohta}}, \bibinfo {author} {\bibfnamefont {Y.}~\bibnamefont {Ota}}, \bibinfo
  {author} {\bibfnamefont {M.}~\bibnamefont {Nomura}}, \bibinfo {author}
  {\bibfnamefont {N.}~\bibnamefont {Kumagai}}, \bibinfo {author} {\bibfnamefont
  {S.}~\bibnamefont {Ishida}}, \bibinfo {author} {\bibfnamefont
  {S.}~\bibnamefont {Iwamoto}}, \ and\ \bibinfo {author} {\bibfnamefont
  {Y.}~\bibnamefont {Arakawa}},\ }\href {\doibase
  http://dx.doi.org/10.1063/1.3579535} {\bibfield  {journal} {\bibinfo
  {journal} {Appl. Phys. Lett.}\ }\textbf {\bibinfo {volume} {98}},\ \bibinfo
  {eid} {173104} (\bibinfo {year} {2011})}\BibitemShut {NoStop}%
\bibitem [{\citenamefont {Schrieffer}\ and\ \citenamefont
  {Wolff}(1966)}]{schrieffer66}%
  \BibitemOpen
  \bibfield  {author} {\bibinfo {author} {\bibfnamefont {J.~R.}\ \bibnamefont
  {Schrieffer}}\ and\ \bibinfo {author} {\bibfnamefont {P.~A.}\ \bibnamefont
  {Wolff}},\ }\href {\doibase 10.1103/PhysRev.149.491} {\bibfield  {journal}
  {\bibinfo  {journal} {Phys. Rev.}\ }\textbf {\bibinfo {volume} {149}},\
  \bibinfo {pages} {491} (\bibinfo {year} {1966})}\BibitemShut {NoStop}%
\bibitem [{\citenamefont {Sekatski}\ \emph {et~al.}(2010)\citenamefont
  {Sekatski}, \citenamefont {Sanguinetti}, \citenamefont {Pomarico},
  \citenamefont {Gisin},\ and\ \citenamefont {Simon}}]{sekatski10}%
  \BibitemOpen
  \bibfield  {author} {\bibinfo {author} {\bibfnamefont {P.}~\bibnamefont
  {Sekatski}}, \bibinfo {author} {\bibfnamefont {B.}~\bibnamefont
  {Sanguinetti}}, \bibinfo {author} {\bibfnamefont {E.}~\bibnamefont
  {Pomarico}}, \bibinfo {author} {\bibfnamefont {N.}~\bibnamefont {Gisin}}, \
  and\ \bibinfo {author} {\bibfnamefont {C.}~\bibnamefont {Simon}},\ }\href
  {\doibase 10.1103/PhysRevA.82.053814} {\bibfield  {journal} {\bibinfo
  {journal} {Phys. Rev. A}\ }\textbf {\bibinfo {volume} {82}},\ \bibinfo
  {pages} {053814} (\bibinfo {year} {2010})}\BibitemShut {NoStop}%
\bibitem [{\citenamefont {Essler}\ \emph {et~al.}(2005)\citenamefont {Essler},
  \citenamefont {Frahm}, \citenamefont {G\"ohmann}, \citenamefont {Kl\"umper},\
  and\ \citenamefont {Korepin}}]{essler05}%
  \BibitemOpen
  \bibfield  {author} {\bibinfo {author} {\bibfnamefont {F.~H.~L.}\
  \bibnamefont {Essler}}, \bibinfo {author} {\bibfnamefont {H.}~\bibnamefont
  {Frahm}}, \bibinfo {author} {\bibfnamefont {F.}~\bibnamefont {G\"ohmann}},
  \bibinfo {author} {\bibfnamefont {A.}~\bibnamefont {Kl\"umper}}, \ and\
  \bibinfo {author} {\bibfnamefont {V.~E.}\ \bibnamefont {Korepin}},\
  }\href@noop {} {\emph {\bibinfo {title} {The One-Dimensional Hubbard
  Model}}}\ (\bibinfo  {publisher} {Cambridge University Press},\ \bibinfo
  {year} {2005})\BibitemShut {NoStop}%
\bibitem [{\citenamefont {Winkler}(2003)}]{winkler03}%
  \BibitemOpen
  \bibfield  {author} {\bibinfo {author} {\bibfnamefont {R.}~\bibnamefont
  {Winkler}},\ }\href@noop {} {\emph {\bibinfo {title} {Spin-Orbit Coupling
  Effects in Two-Dimensional Electron and Hole Systems}}}\ (\bibinfo
  {publisher} {Springer-Verlag},\ \bibinfo {year} {2003})\BibitemShut {NoStop}%
\end{thebibliography}
\end{document}